\documentclass[acmtog, screen]{acmart}

\usepackage{subcaption}
\usepackage{graphicx}
\usepackage{algorithm}
\usepackage{algpseudocode}
\usepackage{multirow}
\usepackage{wrapfig}
\usepackage{pdfpages}
\usepackage{graphicx}
\usepackage{soul}
\algrenewcommand\textproc{}

\newcommand{\mbX}{\ensuremath{\mathbf{X}}}
\newcommand{\mbA}{\ensuremath{\mathbf{A}}}
\newcommand{\mbI}{\ensuremath{\mathbf{I}}}
\newcommand{\mbG}{\ensuremath{\mathbf{G}}}
\newcommand{\mbF}{\ensuremath{\mathbf{F}}}
\newcommand{\mbx}{\ensuremath{\mathbf{x}}}

\newcommand{\secref}[1]{Section~\ref{sec:#1}}

\renewcommand{\eqref}[1]{Equation~\ref{eq:#1}}

\copyrightyear{2026}
\acmYear{2026}
\setcopyright{cc}
\setcctype{by}
\acmConference[SIGGRAPH Conference Papers '26]{Special Interest Group on Computer Graphics and Interactive Techniques Conference Conference Papers}{July 19--23, 2026}{Los Angeles, CA, USA}
\acmBooktitle{Special Interest Group on Computer Graphics and Interactive Techniques Conference Conference Papers (SIGGRAPH Conference Papers '26), July 19--23, 2026, Los Angeles, CA, USA}
\acmDOI{10.1145/3799902.3811199}
\acmISBN{979-8-4007-2554-8/2026/07}

\AtBeginDocument{%
	\providecommand\BibTeX{{%
			\normalfont B\kern-0.5em{\scshape i\kern-0.25em b}\kern-0.8em\TeX}}}

\begin{document}
	
    \title{AGIPC: Adaptive In-Solve Algebraic Coarsening for GPU IPC}
    
    \author{Xuan Wang}
    \orcid{0009-0004-7061-8321}
	\email{xuan-wang@connect.hku.hk}
	\affiliation{%
		\institution{The University of Hong Kong}
		\country{Hong Kong SAR}
	}

    \author{Zhaofeng Luo}
    \orcid{0009-0007-8947-9108}
	\email{zhaofen2@andrew.cmu.edu}
	\affiliation{%
		\institution{Carnegie Mellon University}
		\country{USA}
	}

    \author{Minchen Li}
    \orcid{0000-0001-9868-7311}
	\email{minchernl@gmail.com}
	\affiliation{%
		\institution{Carnegie Mellon University, Genesis AI}
		\country{USA}
	}
        
	\author{Taku Komura}
	\orcid{0000-0002-2729-5860}
	\email{taku@cs.hku.hk}
	\affiliation{%
		\institution{The University of Hong Kong}
		\country{Hong Kong SAR}
	}

    \author{Kemeng Huang}
    \authornote{Corresponding author and project lead}
	\orcid{0000-0001-9147-2289}
	\email{kmhuang@connect.hku.hk}
	\email{kmhuang819@gmail.com}
	\affiliation{%
		\institution{The University of Hong Kong}
		\country{Hong Kong SAR}
	}
        
	\renewcommand{\shortauthors}{Xuan Wang, Zhaofeng Luo, Minchen Li, Taku Komura, Kemeng Huang}

	\begin{abstract}

Implicit time integration is key to robustly simulating stiff materials and large deformations, but its performance is often dominated by repeatedly solving large linear systems. Adaptive coarsening can reduce this cost by concentrating degrees of freedom (DoF) to where it is most needed, yet conventional explicit remeshing changes connectivity (and often vertex ordering), complicating parallel implementations, harming memory locality, and sometimes being disallowed when it may introduce local geometry intersections. Adaptive subspace approaches avoid topological changes, but basis construction and updates incur irregular data access patterns and typically produce dense system matrices, limiting GPU efficiency and keeping many practical systems CPU-centric.
We present \textit{algebraic adaptive in-solve coarsening}, a GPU-oriented method that dynamically reduces DoF within the Newton solve of implicit time integration without explicit topological modification. Starting from a fine mesh, we express adaptivity as a selective edge-collapse process governed by per-edge tags. Collapsible edges are aggregated in parallel using a warp-level hash mapping scheme that groups fine vertices into coarse ``super-nodes'', while protected edges preserve local detail. This defines an implicit coarse mesh whose linear system is assembled algebraically by mapping and reducing fine-scale gradients and Hessians via efficient GPU reduction kernels. We solve the resulting coarse system with a preconditioned conjugate gradient (PCG) method and then prolongate the solution back to the fine mesh. Our approach integrates seamlessly with IPC’s barrier energy and exploits GPU parallelism end-to-end. Across a range of challenging scenarios, we achieve up to 3$\times$ speedup over a state-of-the-art GPU IPC solver while producing visually indistinguishable results. 
	\end{abstract}

	
	
	\begin{teaserfigure}
		\centering
		\begin{subfigure}[b]{0.40\textwidth}
			\centering
			\includegraphics[width=\columnwidth, trim=200 200 200 0, clip]{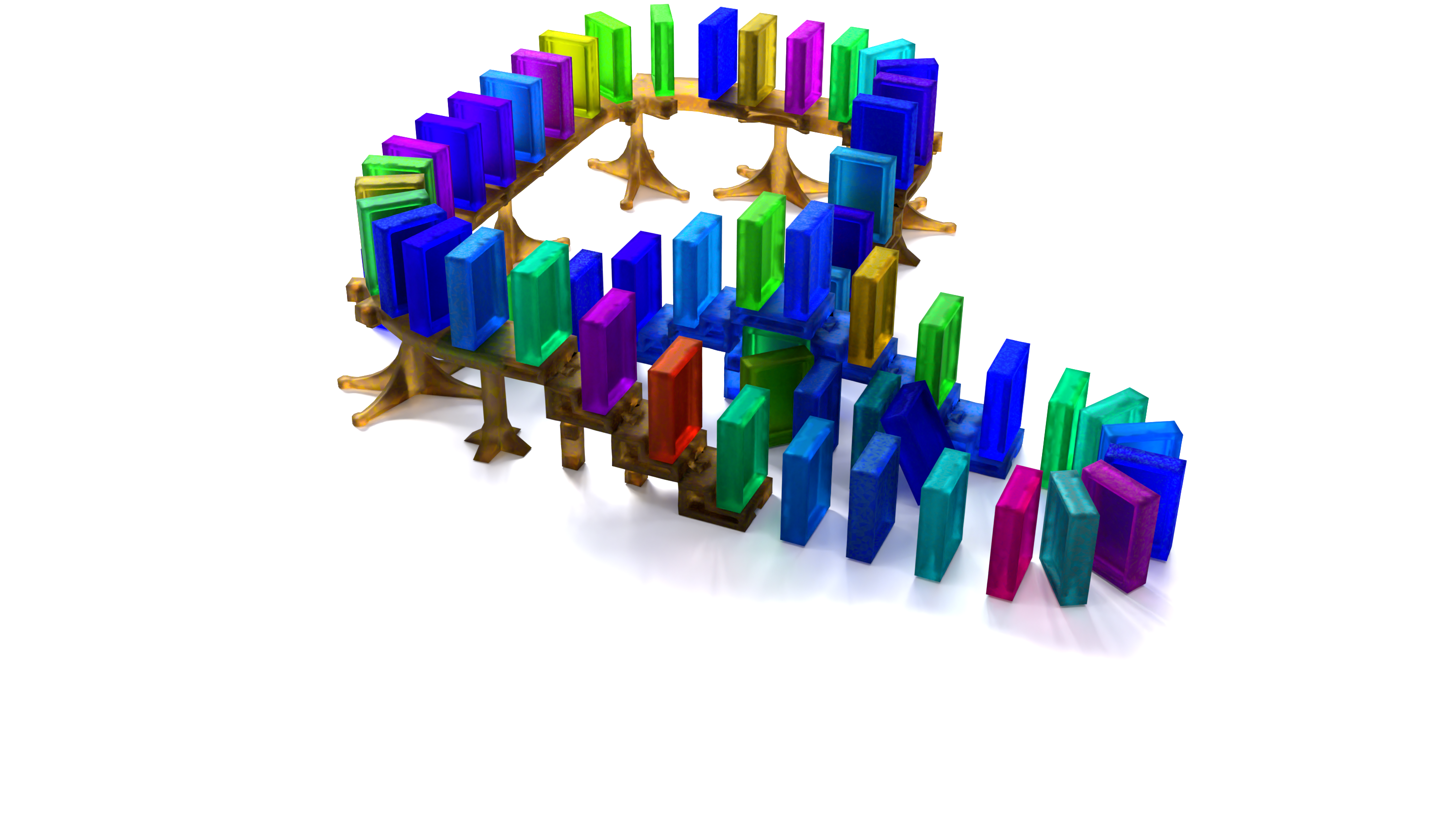}
		\end{subfigure}
        \hspace{1cm}
		\begin{subfigure}[b]{0.40\textwidth}
			\centering
			\includegraphics[width=\columnwidth, trim=200 180 200 20, clip]{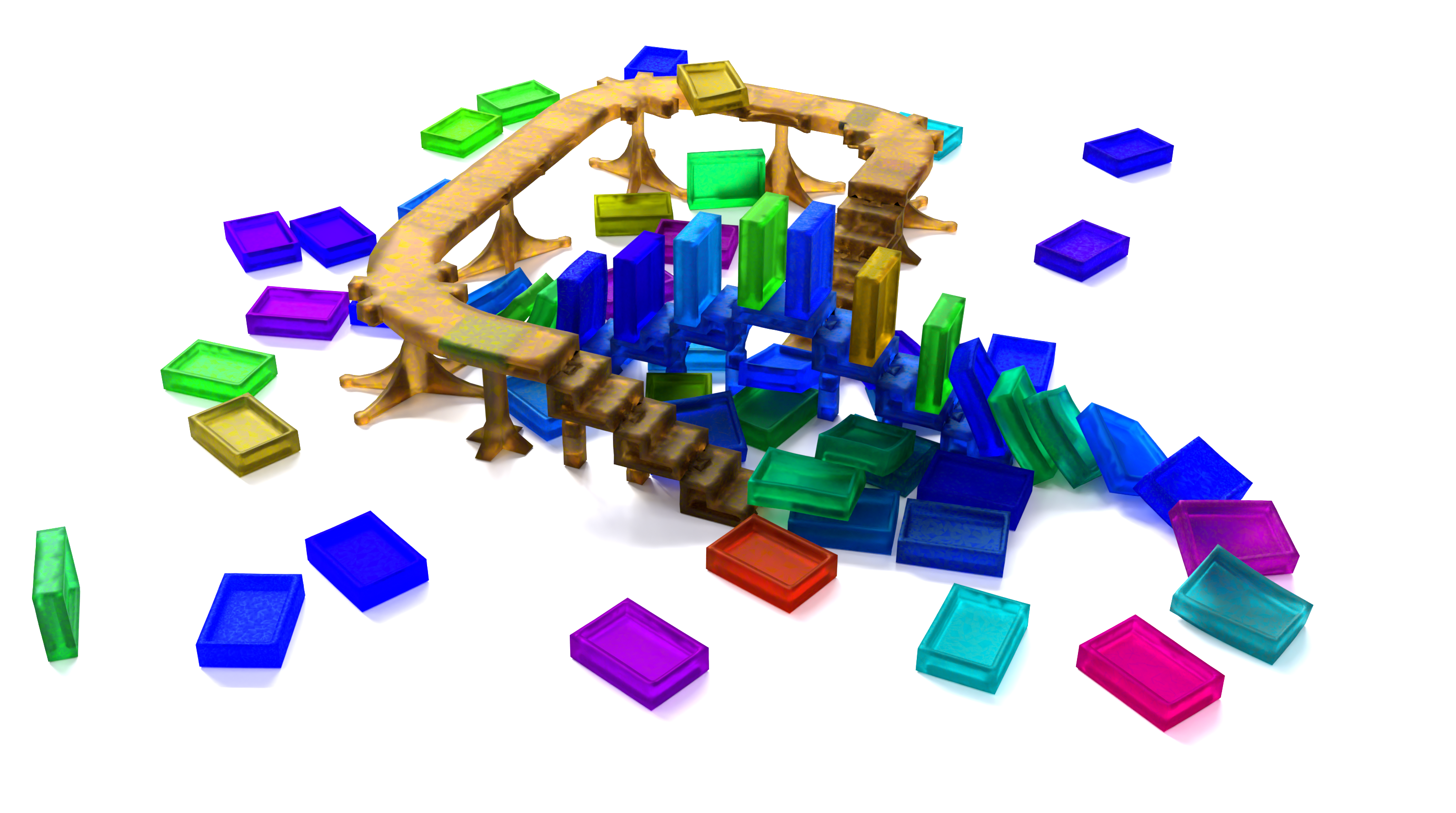}
		\end{subfigure}
		\caption{\textbf{Soft domino.} We simulate a domino scene with deformable cuboids (Young's Modulus is $3\times10^5\text{Pa}$) and rigid holders (simulated with ABD). As the dominoes fall sequentially, the kinetic and elastic energy distribution of the scene evolves over time. Our algebraic coarsening method adaptively coarsens inactive degrees of freedom at each Newton iteration, achieving a $2\times$ speedup over StiffGIPC while producing visually comparable results.}
		\label{fig:teaser}
	\end{teaserfigure}
	
	\maketitle
	
	\section{Introduction}
    Implicit time integration \cite{DBLP:conf/siggraph/BaraffW98} provides essential numerical stability for simulating large deformations of stiff and soft materials.
    When applied in the Incremental Potential Contact (IPC) method \cite{ipc, cipc}, it robustly handles complex frictional contact. However, this stability comes at a high computational cost: each time step requires solving a large, sparse nonlinear system derived from the system energies' Hessian and gradient. Its $O(n^2)$ cost grows with the number of degrees of freedom (DoF) $n$, making high resolution simulations prohibitively expensive.

A promising approach to reduce cost is adaptive remeshing, which dynamically coarsens and refines the simulation mesh based on deformation or error metrics \cite{adaptive-tearing, DBLP:journals/tog/NarainPO13, DBLP:journals/tog/WickeRKBSO10, DBLP:journals/tog/NarainSO12}. While often effective, explicit topological changes disrupt IPC’s barrier energy, violating the $C^2$ continuity needed for reliable convergence in contact-rich scenarios. To preserve an energy descent search direction, remeshing is therefore usually postponed until after the Newton solver has converged for the current time step, followed by the necessary variable remapping \cite{DBLP:journals/tog/FergusonSKP23}. In addition, these remeshing operations introduce irregular memory access patterns that are poorly suited to GPUs.

Recent adaptive subspace method \cite{TrustyFLK24} enables in-solve integration with IPC, but remain tailored for CPUs and inefficient on GPUs. Similar to explicit remeshing, this method relies on explicit dense basis construction and maintenance, which again involves irregular data structures and necessitating conservative, oversized memory pre‑allocation on GPU to guard against runtime reallocation during dynamic updates. These factors hinder efficient GPU parallelization. Consequently, building efficient GPU-accelerated adaptive IPC solvers remains an open challenge.


In this work, we introduce algebraic adaptive coarsening, a GPU-friendly approach that dynamically reduces DoF within each Newton iteration of IPC without explicit mesh modifications. We reframe adaptivity as a selective, algebraic edge collapse process: at each Newton iteration, starting from a fine mesh, we tag edges as collapsible or protected based on their local deformation. A parallel, warp-based hashing scheme then aggregates fine nodes into coarse "super-nodes" on-the-fly, and the corresponding coarse linear system is assembled through efficient GPU reduction kernels that map and accumulate fine mesh gradient and Hessian. This purely algebraic coarsening preserves IPC's barrier continuity and force conservation, yielding a valid descent direction while significantly reducing the size of the linear system.
Our approach is particularly effective in scenes with large-scale coherent motion, where aggressive coarsening incurs minimal error, and generalizes seamlessly among rods, shells, and volumetric meshes.

A summary of our contributions is as follows:
\begin{itemize}
\item A GPU friendly algebraic adaptive coarsening method for large scale simulations that delivers significant performance gains without sacrificing visual fidelity. This is achieved by combining two synergistic components: an algebraic coarsening strategy that efficiently maps the fine scale system to a coarse scale while avoiding costly topological changes and irregular memory access, and an adaptive affine embedding scheme governed by a Green strain increment criterion, which dynamically preserves essential rotational motions and minimizes approximation error. 
\item The first fully GPU optimized adaptive IPC solver{\footnote{\color{black}source code link: https://github.com/KemengHuang/Adaptive-GIPC}}, achieving up to $3\times$ speedup over state-of-the-art GPU IPC framework with visually identical results. We validate our framework through extensive benchmarks against StiffGIPC and alternative design choices, demonstrating superior performance across soft and stiff materials, variable time steps, hybrid simulations, and large-scale scenarios with thousands of objects.
\end{itemize}


	\section{Related Works}

In this section, we provide a brief review of DoF reduction strategies, categorized into three primary categories: remeshing-based approaches that adapt spatial discretization; subspace methods projecting dynamics onto lower-rank manifolds; and multigrid methods utilizing hierarchical coarsening to accelerate solver convergence. We further survey recent GPU-based optimizations for the IPC framework.

\paragraph{Remeshing Methods.}
Remeshing methods \citep{DBLP:journals/cgf/ManteauxWNRFC17} adapt spatial discretization to concentrate DoFs in regions of high geometric or dynamical complexity. Typical remeshing criteria include strain- or stress-based measures for solids \citep{DBLP:journals/tog/WickeRKBSO10}, wrinkle- and contact-aware metrics for cloth \citep{DBLP:journals/tog/NarainSO12, DBLP:journals/tog/NarainPO13, DBLP:journals/tog/LiDNBOBB18}, and elastic or contact energy–based indicators in IPC-style simulations \citep{DBLP:journals/tog/FergusonSKP23, DBLP:journals/tog/WenBK25}. However, these criteria are often material-dependent and require scenario-specific tuning. 
Moreover, robust remeshing operations remain challenging. Global remeshing is computationally expensive \citep{DBLP:journals/tog/KlingnerFCO06, DBLP:journals/tog/BargteilWHT07, DBLP:journals/tog/Grinspun14, DBLP:journals/tog/JiangSP17}, while local updates, such as edge splits and collapses, can cause element inversions and self-intersections. Remedies either use post-processing \citep{DBLP:journals/tog/NarainSO12}, which lacks robustness guarantees, or conservatively reject operations \citep{DBLP:journals/tog/FergusonSKP23}, limiting DoF reduction and potentially degrading mesh quality. Furthermore, discrete connectivity changes can disrupt trajectory consistency and hinder convergence.
In contrast, our method avoids explicit geometric modification by keeping the spatial discretization fixed and performing DoFs reduction algebraically at the linear system level.

{\color{black}
\paragraph{Subspace Methods.}
Subspace methods reduce computational cost by projecting high-dimensional dynamics onto low-rank bases. Classical Hessian-eigenvector approaches \citep{10.1145/74333.74355} are efficient, but their linearity limits performance under large rotations. Subsequent research address this limitation through modal warping \citep{DBLP:journals/tvcg/ChoiK05}, modal derivatives \citep{DBLP:journals/tog/BarbicJ05}, rotation-strain coordinates \citep{DBLP:journals/tog/PanBH15}. Skinning-space methods \citep{DBLP:journals/tog/GillesBFP11, DBLP:journals/tog/JacobsonBPS11, DBLP:journals/tog/Jacobson15, DBLP:journals/tog/BenchekrounZCGZJ23} naturally encode rotation. Linear blend skinning \citep{DBLP:journals/tog/BenchekrounZCGZJ23} represents displacements as weighted sums of affine transformations, inspiring our affine embedding approach. A similar method is also employed in Affinification \citep{mercier2026affinification}, a recent concurrent work.
To preserve fidelity, \citet{DBLP:journals/tog/KimJ09} incrementally build a reduced nonlinear model from full-simulation snapshots and use error estimators to predict its validity. More recently, \citet{TrustyFLK24} integrate subspace reduction within IPC by adaptively activating a subset of precomputed basis functions, augmented with nodal enrichment.
Nevertheless, subspace methods share certain limitations: accuracy is bounded by the expressiveness of the chosen bases. Furthermore, despite efforts to improve locality and sparsity \citep{DBLP:journals/tog/0007VWWMT13}, subspace bases often remain dense due to the global support of vibration modes, leading to memory-intensive coarse Hessian construction. To address this, we implicitly constructs the subspace via a lightweight mapping function, incurring less than $10ms$ of overhead in our practice.
}

{\color{black}
\paragraph{Multigrid Methods.}
Multigrid (MG) methods \cite{DBLP:books/daglib/0002128} achieve optimal linear scaling by hierarchically damping errors: stationary smoothers resolve high-frequency components, while coarse-grid corrections address low-frequency residuals. Although our framework employs a similar fine-to-coarse hierarchy, its design diverges from traditional MG. MG methods typically function as preconditioners or linear solvers, whereas our approach adaptively constructs a reduced-order manifold to bypass the full-space solve entirely. 
Existing MG techniques include geometric multigrid (GMG) \citep{DBLP:journals/cgf/JeonCKCK13, DBLP:journals/tog/XianTL19, DBLP:journals/tog/LuYMH25}, which derives hierarchies from geometric subdivisions. While recent GPU-based GMG offers high throughput \citep{DBLP:journals/tog/LuYMH25}, it lacks the line-search robustness required for IPC. We therefore compare against \citet{DBLP:journals/tog/XianTL19}, which remains influential and shares a similar affine transformation with our method. Alternatively, algebraic multigrid (AMG) \citep{DBLP:journals/tog/TamstorfJM15, DBLP:journals/tvcg/TakahashiB25} constructs hierarchies directly from the coefficient matrix. While AMG offers superior convergence, its setup cost is significant. Many modern libraries (e.g., PETSc \citep{DBLP:conf/scitools/BalayGMS96}, HYPRE \citep{DBLP:conf/iccS/FalgoutY02}) support AMG solvers. We benchmark against NVIDIA's AmgX \citep{DBLP:journals/siamsc/NaumovACCDELMRS15} as a highly optimized GPU baseline.
Despite their effectiveness for elliptic PDEs, MG often struggle with the extreme nonlinearity and localized stiffness inherent in IPC. Nonetheless, multiresolution principles remain a fertile research area, exemplified by multi-layer solvers \citep{DBLP:journals/cgf/MercierAubinK24} and progressive dynamics \citep{DBLP:journals/tog/ZhangJK25}.
}

\paragraph{GPU Optimization for IPC}
Despite the accuracy and robustness of IPC, its high computational cost remains a bottleneck for large-scale simulations. To mitigate this issue, many works exploit GPU parallelism to accelerate IPC-related computations. Early approaches \citep{DBLP:journals/tog/Lan0KYLJ21,DBLP:journals/tog/LanKLJY22} adopt hybrid CPU–GPU pipelines. GIPC \citep{gipc} presents the first fully GPU-based IPC implementation, enabled by an efficient Hessian approximation that avoids the GPU-unfriendly eigendecomposition in the original barrier formulation. Similarly, \citet{DBLP:journals/ral/DuYWXXL24} propose a GPU-based IPC solver for unified soft–rigid body dynamics in robotics. More recently, StiffGIPC \citep{stiffgipc} further accelerates GIPC by introducing a connectivity-aware multilevel additive Schwarz (MAS) preconditioner \citep{wu2022gpu} to improve PCG convergence, along with a hash-based two-level reduction strategy for efficient Hessian assembly. While GPUs are well suited for dense, regular, and massively parallel workloads, adaptive techniques such as remeshing \citep{DBLP:journals/tog/FergusonSKP23} and adaptive subspace methods \citep{TrustyFLK24} introduce irregular memory access, dynamic data structures, and frequent synchronization, which significantly hinder efficient GPU utilization.

\begin{figure*}[htbp]
\centering
\begin{subfigure}[b]{0.95\textwidth}
\includegraphics[width=\columnwidth, trim=0 0 0 0, clip]{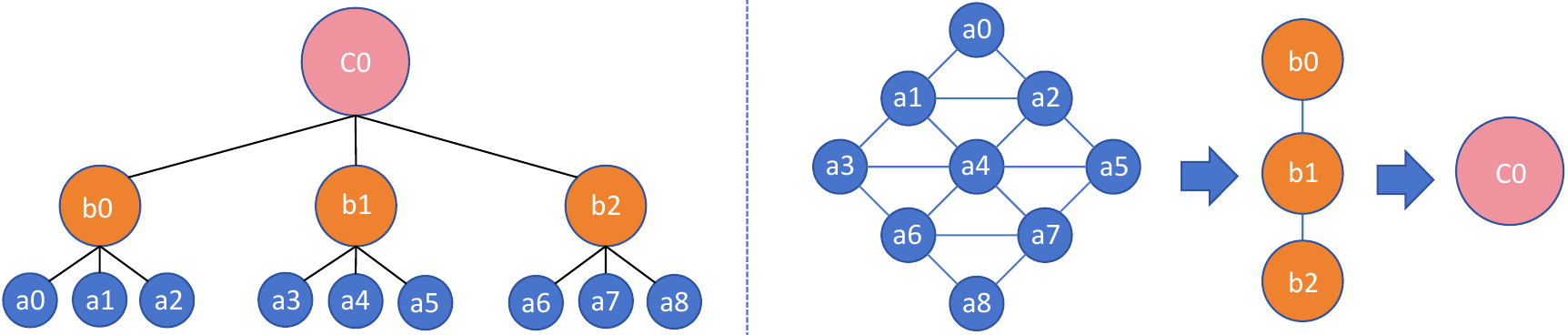}
\end{subfigure}
\caption{
\textbf{A fine mesh coarsening example.} In the left figure, three consecutive nodes are grouped. Blue disks are mesh nodes; blue lines (right) are their edges. Fully connected nodes within a group are aggregated into a single, coarser node (orange disks). This grouping and aggregation are applied recursively. The final coarsened mesh is shown as a large pink disk, the coarse supernode, achieving the target minimum DoF.
}
\label{fig:remesh_init}
\end{figure*}

\begin{algorithm}
    \caption{Algorithm overview of adaptive GPU IPC}
    \begin{algorithmic}[1]
        \vspace{0.1cm}
        \Statex{\textbf{init()}}
        
        \For{each time step $t$} 
            \State $\mathbf{x} \leftarrow \mathbf{x}^t$ 
            \Repeat
                \State $E_{prev} \leftarrow E(\mathbf{x})$ 
                \State $\mathbf{x}_{prev} \leftarrow \mathbf{x}$ 
                \State $\nabla E(\mathbf{x}) \leftarrow \text{update fine mesh gradient}$
                \State $\mathbf{H}(\mathbf{x}) \leftarrow \text{update fine mesh Hessian}$
                \State $map \leftarrow \text{update fine-coarse mesh mapping function}$
                \State $\mathbf{H}_c, \mathbf{g}_c \leftarrow map({\mathbf{H}(\mathbf{x}), \nabla E(\mathbf{x})})$
                \State $\mathbf{d}_c = -{\mathbf{H}_c}^{-1}\mathbf{g}_c$
                \State $\mathbf{d} \leftarrow \text{post\_coarsening}(\mathbf{d}_c, \mathbf{H}(\mathbf{x}), \nabla E(\mathbf{x}))$
                \State update\_BVH($\mathbf{x}$)
                \State find\_CCD\_potential\_collision\_pairs($\mathbf{x}$)
                \State $\alpha \leftarrow\min(1, \mathrm{maxFeasibleStepSize})$
                \Repeat
                    \State $\mathbf{x} \leftarrow \mathbf{x}_{prev} + \alpha \mathbf{d}$
                    \State update\_BVH($\mathbf{x}$)
                    \State find\_CD\_collision\_pairs($\mathbf{x}$)
                    \State $\alpha = 0.5*\alpha$
                \Until{$\text{$E(\mathbf{x})$<$E_{prev}$}$} 
            \Until{$\frac{\|\mathbf{d}\|_{\infty}}{\Delta{t}} \le \varepsilon_d$} 
        \EndFor
    \end{algorithmic}\label{alg:overview}
\end{algorithm}

    \section{Background and preliminaries}
	\label{sec:background}
        IPC~\cite{ipc} formulates implicit time integration for elastodynamic contact as the minimization of an Incremental Potential (IP):
    \begin{equation}\label{eq:E_s}
	E(\mbx) = \frac{1}{2}(\mathbf{x} -\mathbf{\hat{\mathbf{x}}})^T\mathbf{M}(\mathbf{x} -\mathbf{\hat{\mathbf{x}}})+\Delta t^2\Psi(\mathbf{x})+B(\mathbf{x})+D(\mathbf{x})
	\end{equation}
    where $\mathbf{x}$ is node positions, and $\hat{\mathbf{x}} = \mathbf{x}^t+\Delta t \mathbf{v}^t+\Delta t^2 \mathbf{M}^{-1} \mathbf{f}_e$ denotes the predicted node positions, $\mathbf{M}$ is the mass matrix, $\mathbf{f}_e$ is the external force, $B$ is the contact barrier potential, and $D$ is the approximated friction potential. Velocities are updated as $\mathbf{v}^{t+\Delta{t}} = (\mathbf{x}^{t+\Delta{t}}-\mathbf{x}^t)/\Delta t$.

    To minimize $E(\mathbf{x})$, a line search method is used. In each iteration $i$, a quadratic proxy is built:
	\begin{equation}\label{eq:tailor}
	E_i(\mathbf{x}) = E(\mathbf{x}_i) + \left(\mathbf{x} - \mathbf{x}_i\right)^T\nabla E(\mathbf{x}_i) + \frac{1}{2}\left(\mathbf{x} - \mathbf{x}_i\right)^T \mathbf{H}(\mathbf{x}_i) \left(\mathbf{x} - \mathbf{x}_i\right),
	\end{equation}
    The descent direction $\mathbf{d}$ is obtained by solving
    $\mathbf{H}(\mathbf{x}_i)  \mathbf{d} = -\nabla E(\mathbf{x}_i)$, where $\mathbf{H}$ is a symmetric positive definite (SPD) proxy matrix approximating $\nabla^2 E$ for fast convergence. The new iterate is computed as $\mathbf{x}_{i+1} = \mathbf{x}_i + \alpha\mathbf{d}$, with $\alpha$ determined via backtracking line search and filtered to maintain contact feasibility. Minimization stops when $\frac{\|\mathbf{d}\|_\infty}{\Delta t} \le \varepsilon_d$, where $\varepsilon_d$ is the Newton tolerance.

    For efficient GPU solving of $\mathbf{H}(\mathbf{x}_i)  \mathbf{d} = -\nabla E(\mathbf{x}_i)$, Stiff-GIPC \cite{stiffgipc} proposes a fast preconditioned conjugate gradient (PCG) solver by customizing a connectivity-enhanced Multilevel Additive Schwartz (MAS) \cite{wu2022gpu} preconditioner and sparse matrix vector (SpMV) multiplication. Their method improves PCG convergence with reduced per iteration cost, and maintain high performance across varying stiffnesses. However, the $O(n^2)$ scaling of Hessian matrix related costs still significantly limit performance as degrees of freedom $n$ increase. Reducing $n$ while preserving simulation accuracy is therefore essential. To mitigate this issue, we now introduce our algebraic adaptive coarsening method, and the overall procedure is summarized in Algorithm \ref{alg:overview}.

\section{Algebraic Adaptive Coarsening}

We present an algebraic adaptive coarsening method designed for robust, high performance simulations with frictional contact, specifically within the GPU IPC framework. Traditional remeshing alters mesh topology through explicit geometric operations (edge splits, collapses), which introduce discontinuities in IPC's $C^2$-continuous barrier energies and can cause solver failure. Instead, our method begins with a fine, static mesh and achieves adaptivity through algebraic coarsening: fine DoF are aggregated into coarse "super nodes" based on edge collapse decisions, and the corresponding linear system (gradient and Hessian) is constructed via efficient parallel reductions (lines 6-9, Algorithm \autoref{alg:overview}). The coarse solution is then mapped back to the fine mesh (lines 10-11, Algorithm \autoref{alg:overview}). This process is performed algebraically each Newton step, avoiding explicit topological changes, preserving barrier continuity, and maintaining a GPU-friendly data layout
(see \secref{coarsening}). 
The adaptivity is driven by a kinematic criterion (Green strain increment, see \secref{collapse_criterion}) to preserve detail where deformation is non-uniform, and an optional affine embedding enriches coarse kinematics. 
A final post-coarsening and full CG iterations are conducted to reconstruct high-frequency details (see \secref{postcoarsening}).
The result is a significant reduction in solved DoF while maintaining simulation robustness and dynamic fidelity.

\subsection{Edge Collapse and Algebraic Coarsening}
\label{sec:coarsening}
\begin{figure}[htbp]
\centering
\begin{subfigure}[b]{0.45\textwidth}
\includegraphics[width=\columnwidth, trim=0 0 0 0, clip]{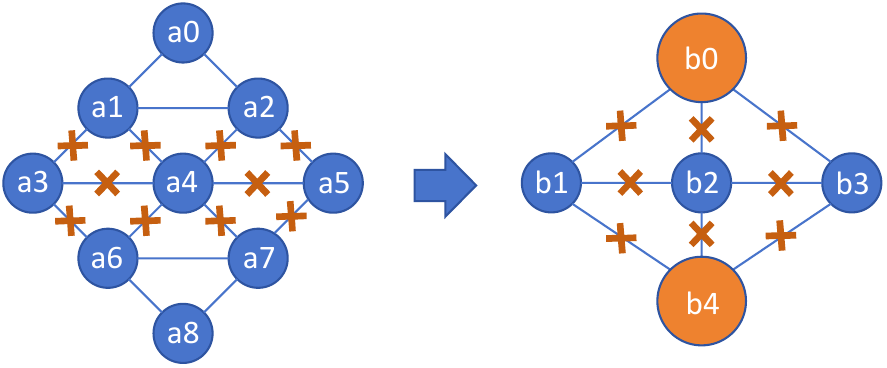}
\end{subfigure}
\caption{
\textbf{A mesh DoF recovery example.} The red-crossed edge indicates that this edge is protected from collapse during the mesh coarsening process.
}
\label{fig:remesh_in_time}
\end{figure}

\paragraph{Adaptive Coarsening via Edge Collapse}
The core of our adaptivity is the selective, algebraic collapse of edges in the fine mesh (see \autoref{fig:remesh_init}). Rather than modifying mesh connectivity, each edge $e$ is assigned a binary tag $\tau_e \in \{0,1\}$. A tag of 1 marks the edge as collapsible; a tag of 0 protects it, preserving local detail (see \autoref{fig:remesh_in_time}). By default, all edges are tagged 1. When a local deformation criterion (detailed in section \ref{sec:collapse_criterion}) is met, the tags of relevant edges are set to 0. A parallel aggregation algorithm then coalesces fine nodes into coarse supernodes, but only across paths consisting of edges tagged 1. This creates a coarse mesh whose resolution varies spatially according to the tag pattern. The process is repeated every Newton iteration, allowing the adaptive resolution to evolve with the simulation. The key advantage is that the fine mesh connectivity remains static; only the mapping from fine to coarse DoF changes. This eliminates topological discontinuities and provides a predictable memory layout for efficient GPU computation. Our parallel aggregation scheme for building the fine-to-coarse mapping is detailed in the supplementary document. 

\paragraph{Algebraic Coarsening and System Assembly}
Once the fine-to-coarse mapping is established, we construct the linear system for the coarse mesh algebraically, without explicitly computing coarse node positions. For the gradient (force vector), the contribution of each fine node $\mathbf{g}_f$ is accumulated to its mapped coarse node $\mathbf{g}_c$: 
 $\mathbf{g}_c = \sum_{f \in child(c)} \mathbf{g}_f$
. The Hessian is assembled similarly: each $3\times3$ fine-level Hessian block, stored in BCOO format with row/column indices $(i, j)$, is mapped to the coarse indices $(map(i), map(j))$. Entries mapping to the same coarse block are summed via a parallel hash reduction \cite{stiffgipc}. This approach is mathematically equivalent to a Galerkin projection $\mathbf{H}_c = \mathbf{U} \mathbf{H}_f \mathbf{U}^T$ and $\mathbf{g}_c = \mathbf{U} \mathbf{g}_f$, where $\mathbf{U}$ is a restriction operator defined by the aggregation mapping. By leveraging efficient GPU parallel primitives (e.g., segmented reduction), the assembly can be done fast; also we can avoid the complex procedure of maintaining explicit coarse meshes \cite{DBLP:journals/tog/FergusonSKP23} (as in traditional remeshing) or dynamic subspace basis \cite{TrustyFLK24} (as in adaptive subspace methods). Additional implementation details of the parallel hash reduction method are provided in the supplementary document. 

\subsection{Adaptive Criteria and Affine Embedding}
\begin{figure}[t]
\centering
\includegraphics[width=\columnwidth]{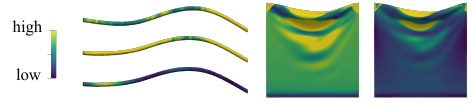}
\caption{
\textbf{Adaptive coarsening criterion.} We use Green strain increments to drive collapse decisions, capturing high-frequency elastic waves in both triangle and tetrahedral meshes. (Left) Initial three Newton iterations of a vibrating string; (Right) first two iterations of a hanging cloth. Because deformation states evolve significantly even within a single Newton step, we re-evaluate our criterion per iteration to maintain simulation fidelity.
}
\label{fig:green_strain}
\end{figure}

\paragraph{Adaptive Coarsening Criterion}
\label{sec:collapse_criterion}
Deciding which edges to protect from collapse is crucial for accuracy. We base this decision on kinematic coherence, using the in-solve Green strain increment $\Delta{\mbG}$ as a local measure of deformation intensity \cite{arigid}. For each element (triangle, tetrahedron, or edge), we compute the Frobenius norm of its strain increment $\|\Delta{\mbG}\|_F$ from the current and previous Newton iterations:
\begin{equation}\label{eq:green_strain_rate}
\Delta{\mbG} = {\mbG_i-\mbG_{i-1}},
\end{equation}
where $\mbG = \frac{1}{2}(\mbF^T\mbF-\mbI)$, $\mbF$ is the element deformation gradient and $\mbI$ is the identity matrix. If the strain increment in any element adjacent to an edge exceeds a threshold, that edge is tagged 0 (protected); otherwise, it remains tag 1 (collapsible). This criterion naturally preserves resolution in regions undergoing rapid deformation or high‑frequency elastic waves, while aggressively coarsening areas with near‑rigid or quasi‑static motion. Since elastic waves can travel large distances even within a single Newton iteration (see \autoref{fig:green_strain}), we update the criterion at each Newton iteration $i$. Because the fine mesh connectivity is static, the edge–element adjacency can be precomputed, making the tagging step efficient and applicable to various element types (shells, volumes, rods).

\begin{figure}[htbp]
\centering
\includegraphics[width=1.0\columnwidth]{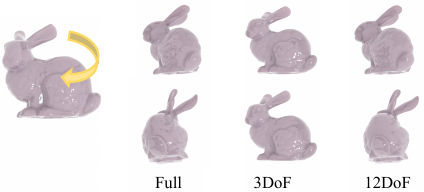}
\caption{
\textbf{Rotating bunny.} A bunny model is undergoing free fall with an initial rotational velocity. Coarsening with only 3 DoF per node fails to capture rotational motion, leading to significant angular momentum loss. By using 12-DoF affine transformations, our method effectively preserves rotation, matching the dynamics of the full-space simulation.
}
\label{fig:rotation_bunny}
\end{figure}

\paragraph{Adaptive Affine Embedding}
Our aggregation that assigns only 3 translational DoF to each coarse node cannot capture relative rotations within an aggregate, leading to artificial stiffening and loss of angular momentum (see \autoref{fig:rotation_bunny}, middle). To mitigate this, we enrich coarse nodes that aggregate many fine nodes with affine DoF. Specifically, for a coarse node $c$ that maps a set of fine nodes $f$, we define a local affine basis $\mbA_f = \bar{\mbX}_f \otimes \mbI_3$, where $\bar{\mbX}_f$ is the homogeneous rest‑pose coordinate of node $f$. The coarse gradient and Hessian are then constructed as
\begin{equation}\label{eq:affine_embedding}
\mathbf{g}_c = \sum_f \mbA_f \mathbf{g}_f \quad \text{and} \quad \mathbf{H}_c = \sum_f \mbA_f \mathbf{H}_f \mbA_f^T,
\end{equation}
giving the coarse node 12 DoF (translation, rotation, scaling, shearing). To balance cost and accuracy, we apply affine embedding only when the number of fine nodes in an aggregate exceeds a threshold (empirically set to 32). For smaller aggregates, we use the standard 3‑DoF mapping. More implementation details of the affine mapping are provided in the supplemental document. 


\subsection{Post-Coarsening and Contact Handling}
\label{sec:postcoarsening}
\paragraph{Post‑Coarsening Refinement and Solution Prolongation}
After solving the reduced coarse system $\mathbf{H}_c \mathbf{d}_c = -\mathbf{g}_c$, we mathematically prolongate the displacement to the fine mesh using the transpose of the restriction operator: $\mathbf{d}_f = \mathbf{U}^T \mathbf{d}_c$. This choice preserves symmetry and is consistent with the Galerkin projection. As the coarse solution may not capture all high frequency details, we perform a small number of conjugate gradient iterations (post coarsening) on the full fine mesh system, starting from $\mathbf{d}_f$. This step acts like a multigrid smoother, correcting local errors and helping to anticipate deformation in subsequent iterations. In practice, fewer than 10 CG iterations suffice, which we adopt as the maximum allowed in our implementation. Finally, contact forces are handled naturally: contact detection and barrier evaluations are performed on the full fine mesh, ensuring $C^2$ continuity. The resulting contact gradients and Hessians are assembled into the fine system before algebraic coarsening, so all contact constraints are preserved in the reduced system.

    \section{Experiments}

We report experimental results in this section. All experiments were conducted on a workstation equipped with an Intel Core i9-14900K CPU (32 cores), 64 GB of RAM, and an NVIDIA RTX 4090 GPU with 24 GB of memory. For all examples, the Newton solver’s tolerance $\varepsilon_d$ is set to $10^{-3}l$ (in m/s), where $l$ is the diagonal length of the scene bounding box, and the PCG solver used a relative residual-norm tolerance of $10^{-3}$. {\color{black}
StiffGIPC terminates when the full-space displacement norm is below tolerance; our method terminates when the prolongated coarse level displacement norm (refined via post-coarsening correction) is below tolerance. 
} {\color{black}Thus, the two frameworks compared in the benchmarks are solving with different accuracy. While not a strict side‑by‑side comparison, this evaluation remains meaningful given the nearly identical visual results.}
All simulations were performed using double-precision floating-point arithmetic. 

\subsection{Comparison and Ablation Study}

\begin{figure}[htbp]
	\centering
	\includegraphics[width=1.0\columnwidth]{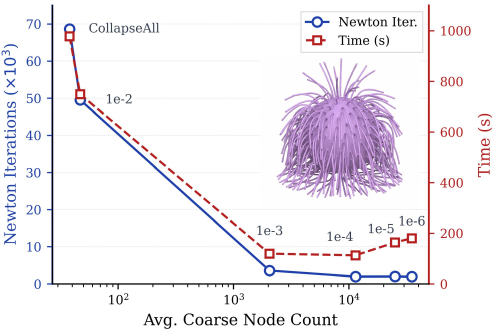}
	\caption{
		\textbf{Varying criterion thresholds.}
		A squishy ball falling onto the ground simulated with different thresholds for the adaptive coarsening criterion. Smaller thresholds preserve finer deformation details but restrict coarsening; beyond a point, insufficient coarsening leads to increased computational cost. Larger thresholds result in more aggressive coarsening; however, due to the post-coarsening step, visually plausible dynamics are still achieved, albeit with increased Newton iterations. Even in the extreme case where all edges are collapsed (purple ball, top right), the resulting motion remains comparable.
	}
	\label{fig:criterion_threshold}
\end{figure}

\paragraph{Coarsening Thresholds.}
The threshold used in our adaptive coarsening criterion has a significant impact on both performance and accuracy. We evaluate a range of thresholds for the Green strain increment criterion, from $10^{-6}$ to $10^{-2}$, including the extreme case where all edges are collapsed. Results are shown in \autoref{fig:criterion_threshold}. Smaller thresholds preserve finer deformation details but restrict coarsening; beyond a certain point, insufficient coarsening leads to a larger number of active degrees of freedom and increased computational cost. In contrast, larger thresholds promote more aggressive coarsening and reduce spatial resolution. Nevertheless, due to the {\color{black}post-coarsening} step, visually plausible dynamics can still be achieved even under loose thresholds, albeit with increased Newton iterations. Based on this trade-off, we select a threshold of $5\times10^{-5}$, which provides a good balance between efficiency and visual fidelity. This value is used in all experiments unless otherwise noted.

\begin{figure}[htbp]
	\centering
	\begin{subfigure}[b]{0.49\columnwidth}
		\centering
		\includegraphics[width=\columnwidth]{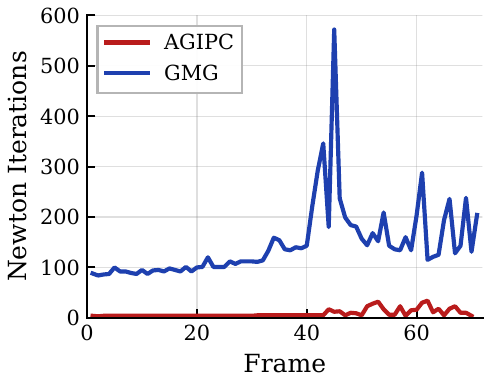}
	\end{subfigure}
	\hfill
	\begin{subfigure}[b]{0.49\columnwidth}
		\centering
		\includegraphics[width=\columnwidth]{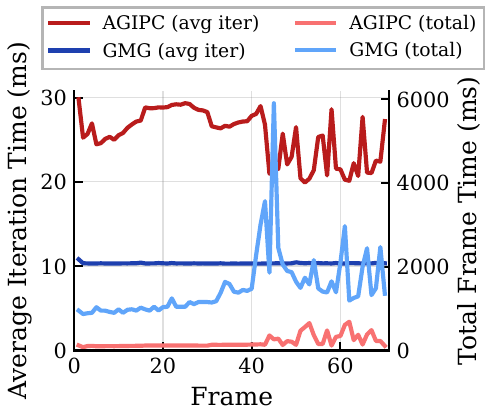}
	\end{subfigure}
	\caption{
		\textbf{Comparison with GMG.}
		Framework-level comparison with a geometric multigrid solver under the same Newton tolerance. GMG requires substantially more Newton iterations to converge. Although individual linear solves are cheaper, the total simulation time for 60 frames is $7.5\times$ longer than ours.
	}
	\label{fig:compare_gmg}
\end{figure}

\begin{figure}[htbp]
	\centering
	\begin{subfigure}[b]{0.49\columnwidth}
		\centering
		\includegraphics[width=\columnwidth]{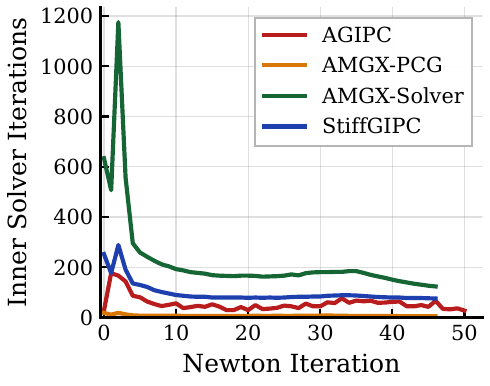}
	\end{subfigure}
	\hfill
	\begin{subfigure}[b]{0.49\columnwidth}
		\centering
		\includegraphics[width=\columnwidth]{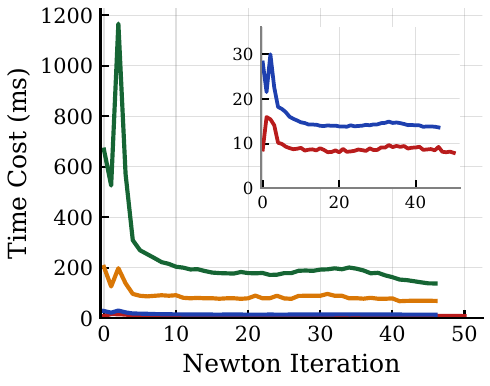}
	\end{subfigure}
	\caption{
		\textbf{Comparison with AmgX.}
		Solver-level comparison with AmgX under identical convergence criteria. When used as a standalone solver (AmgX-Solver), AmgX converges slowly. When used as a PCG preconditioner (AmgX-PCG), it significantly improves convergence; however, its high construction costs result in slower overall performance than both StiffGIPC and our method. A detailed analysis of the V-cycle settings for AmgX is provided in the supplemental document.
	}
	\label{fig:compare_amgx}
\end{figure}

\paragraph{Comparison with Multigrid Methods.}
From the perspective of linear solvers, our method is conceptually related to multigrid approaches. Classical multigrid methods use smoothers to eliminate high-frequency errors on fine levels, while low-frequency errors are addressed on coarser, aggregated levels. Our adaptive coarsening strategy shares similar principle of separating errors by frequency. In our formulation, edges with large Green strain increment (\autoref{eq:green_strain_rate}) exhibit kinematic incoherence, which serves as an indicator of high-frequency error components. Preventing such edges from being collapsed preserves these high-frequency errors on finer levels. Conversely, edges associated with low-frequency errors are preferentially collapsed to coarser levels. Based on this observation, we compare our method with two representative multigrid solvers: a geometric multigrid (GMG) method by \citet{DBLP:journals/tog/XianTL19} and the algebraic multigrid solver AmgX~\citep{DBLP:journals/siamsc/NaumovACCDELMRS15}. We evaluate all methods on the armadillo scene from \citet{DBLP:journals/tog/XianTL19}, which consists of a hanging armadillo model with 40K vertices. We use the StVK elastic energy with Young’s modulus $5\times10^{5}$ and Poisson’s ratio $0.45$. Results are shown in \autoref{fig:compare_gmg} and \autoref{fig:compare_amgx}.

For GMG, we perform a framework-level comparison by running line-search Newton iterations with the same convergence tolerance of $10^{-2}l$. We modify their solver using CUDA~12 with double precision and adopt the two-level multigrid configuration described in their paper. Under identical Newton tolerances, GMG requires substantially more Newton iterations to converge than our method. This behavior is expected, as the fixed geometric hierarchy lacks the adaptivity needed to account for different deformation. Although each GMG linear solve is cheaper on average, the increased number of Newton iterations results in a significantly longer overall simulation time.

    \begin{figure}[htbp]
    	\centering
    	\includegraphics[width=1\columnwidth, trim=0 0 0 0, clip]{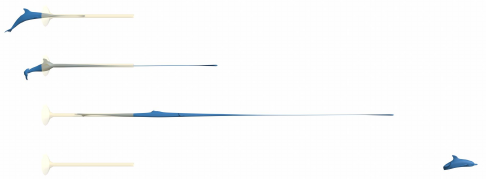}
    	\caption{
    		\textbf{Funnel dolphin.} A soft dolphin is pulled through a thin funnel.
    	}
    	\label{fig:dolphin}
    \end{figure}

For AmgX, we conduct a solver-level comparison replacing our adaptive linear solver with AmgX within the same framework. We choose the \texttt{AGGREGATION} algorithm with the \texttt{MULTICOLOR\_GS} smoother (see the supplemental material for more configuration details). When used directly as a standalone linear solver, AmgX exhibits slow convergence, requiring many iterations to reach the prescribed tolerance. When employed as a preconditioner for PCG, however, AmgX significantly improves convergence, reducing the number of PCG iterations even beyond MAS. Nevertheless, due to its substantial setup cost per Newton iteration and higher preconditioning overhead per iteration, AmgX remains slower overall than both StiffGIPC and our method.

    \begin{figure}[htbp]
    	\centering
    	\includegraphics[width=1\columnwidth, trim=0 0 0 0, clip]{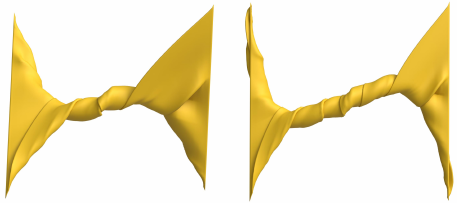}
    	\caption{
    		\textbf{Mat twist.} A stiff mat with Young's modulus $10^{7}$ is twisted by 3 full turns. We set the IPC relative $\hat{d}$ to $3 \times 10^{-4}$ in this scene. We also use a looser coarsening threshold of $5 \times 10^{-4}$ to reduce numerical disturbances under this extreme deformation and high-stiffness regime.
    	}
    	\label{fig:twist}
    \end{figure}

    \begin{figure}[htbp]
	\centering
	\includegraphics[width=1.0\columnwidth]{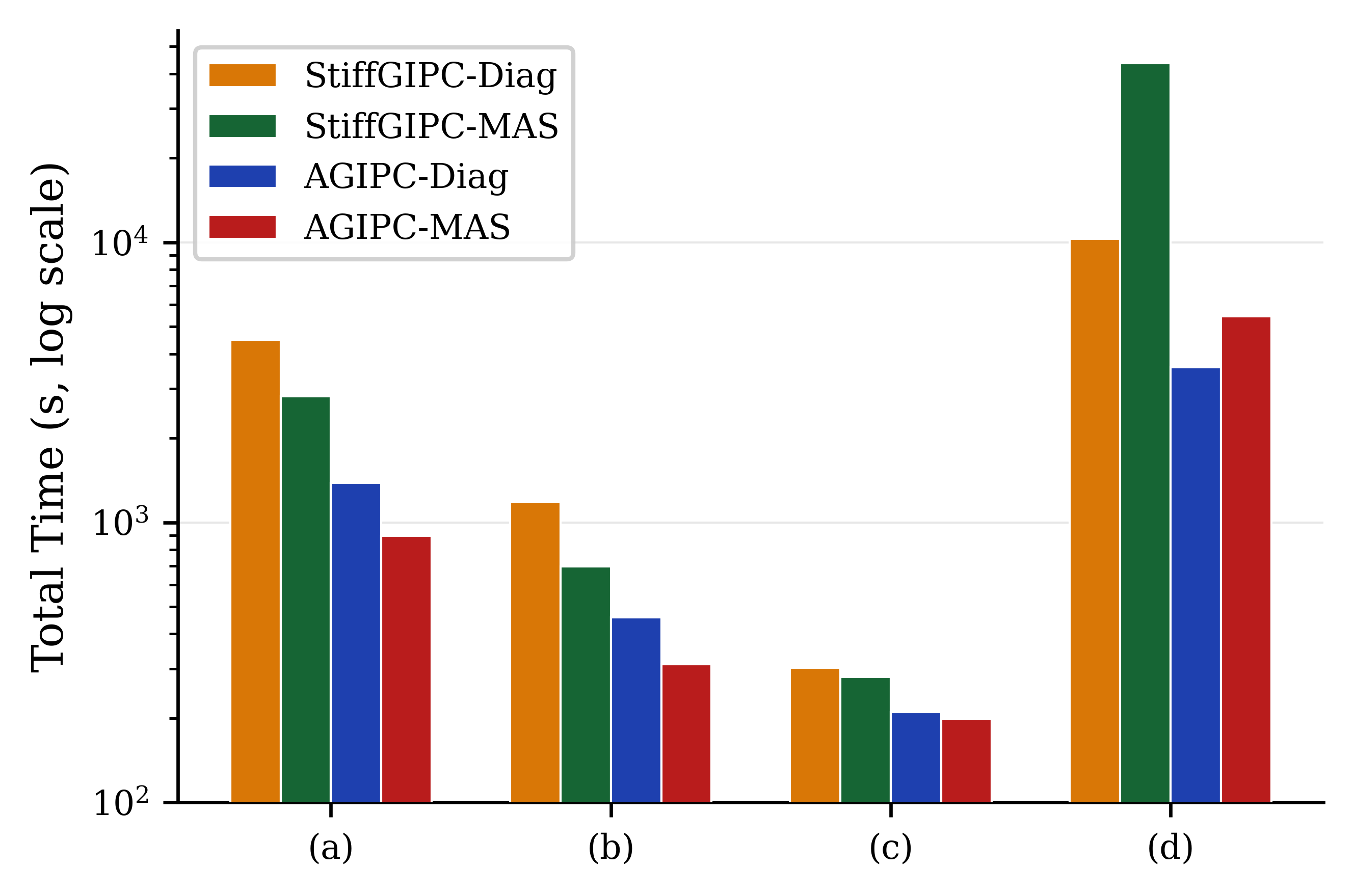}
	\caption{
		\textbf{Effect of preconditioners.}
		Performance comparisons of our method and StiffGIPC using diagonal and MAS preconditioners. (a) Three squishy balls (\autoref{fig:balls_and_xmas} left). (b) Xmas (\autoref{fig:balls_and_xmas} right). (c) Dolphin (\autoref{fig:dolphin}). (d) Twist (\autoref{fig:twist}). MAS typically improves PCG convergence and reduces runtime. But under the extreme deformation and high-stiffness scenario in (d), it can be unstable and slower than the diagonal preconditioner.
	}
	\label{fig:preconditioner}
\end{figure}

\paragraph{Preconditioners.}
The $3\times3$ block Jacobi diagonal preconditioner is widely used for GPU-based PCG solvers due to its simplicity and low overhead. StiffGIPC further introduces an enhanced multilevel additive Schwarz (MAS) preconditioner that leverages mesh connectivity to substantially accelerate PCG convergence. We compare AGIPC and StiffGIPC under both diagonal and MAS preconditioners in \autoref{fig:preconditioner}.
 In most scenarios, the MAS preconditioner effectively reduces the total simulation time, {\color{black} except for the twisting case. A more detailed analysis is provided in the supplemental document.}

    \begin{figure}[htbp]
    \centering
    \includegraphics[width=\columnwidth]{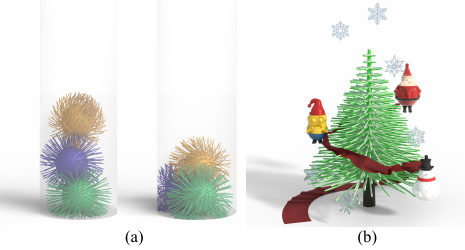}
    \caption{
        (a) \textbf{Three squishy balls.} Three squishy balls with Young's modulus $10^{7}$, $5 \times 10^{6}$, and $10^{6}$, respectively, falling inside a cylinder. (b) \textbf{Xmas.} A Christmas scene featuring a tree with Young's modulus $10^{8}$ and toys with Young's modulus $10^{7}$. The ribbon is a triangular mesh, while the snowflakes are ABD. We set the IPC relative distance $\hat{d}$ to $3e^{-4}$ in both scenes.
    }
    \label{fig:balls_and_xmas}
    \end{figure}


\subsection{Performance Evaluation}

{\color{black}
We compare our method with StiffGIPC across diverse scenarios. Both frameworks use the MAS preconditioner in subsequent experiments, where it outperforms the diagonal preconditioner in all those cases.

}

\begin{figure}[htbp]
	\centering
	\includegraphics[width=1.0\columnwidth]{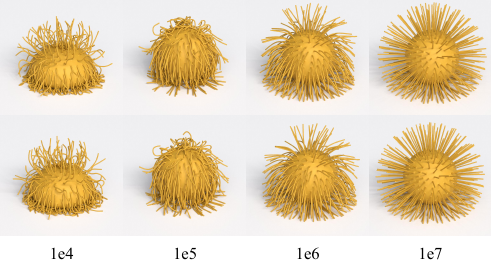}
	\caption{
		\textbf{Varying material stiffness.}
		A squishy ball falling onto the ground simulated with different Young’s modulus. The top row shows full-resolution results, while the bottom row shows results produced by our adaptive method. The dynamics are visually comparable across all stiffnesses. For the stiffest material, our method achieves over $3\times$ speedup compared to StiffGIPC.
	}
	\label{fig:stiffness}
\end{figure}

\paragraph{Varying Material Stiffness.}
We evaluate our method on a range of material stiffness by dropping a squishy ball with 61K vertices and Young’s modulus from $10^{4}$ to $10^{7}$ onto the ground. As stiffness increases, our method achieves progressively larger speedups. For the stiffest case with Young’s modulus $10^{7}$, we obtain $3.30\times$ speedup compared to StiffGIPC \cite{stiffgipc}, while producing visually indistinguishable results (see \autoref{fig:stiffness}). This behavior arises because stiffer materials undergo smaller deformations that quickly damps away, resulting in fewer active degrees of freedom over time and space. Our adaptive coarsening effectively exploits this sparsity, yielding greater computational savings. In contrast, softer materials exhibit larger, long-lasting, and more widespread deformations, which limit coarsening opportunities and reduce the achievable speedup. Our method consistently produces results that closely match the full-resolution simulation across the entire range of stiffness values (see in supplementary video).


    \begin{figure}[htbp]
	\centering
	\includegraphics[width=1\columnwidth]{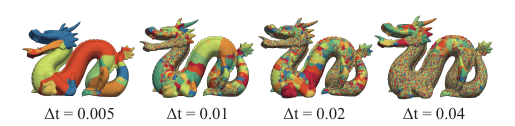}
	\caption{
		\textbf{Visualization of adaptive coarsening under different time step sizes.}
		The mesh is colored according to the vertex mapping from fine to coarse resolution; fine vertices mapped to the same coarse node share the same color. We compute the average number of coarse nodes per Newton iteration, and visualize the coarsening pattern from a representative Newton iteration whose number of coarse nodes most closely matches the per-iteration average. 
	}
	\label{fig:dt}
\end{figure}

\paragraph{Varying Time Step Size.}
We evaluate the behavior of our method under different time step sizes by simulating a dragon model falling onto the ground for a total duration of 1.5\,s. As the time step size $\Delta t$ decreases, we observe a reduction in the average number of active degrees of freedom per Newton iteration. The corresponding coarsening patterns are visualized in \autoref{fig:dt}, where fine vertices mapped to the same coarse node are shown with the same color. From the relevant entries in \autoref{tab:timecost}, we observe that smaller time steps require fewer PCG iterations on average per Newton iteration. 
This trend can be attributed to the fact that smaller time steps limit the amount of deformation that can occur within a single step, resulting in more localized and gradual state changes within Newton iterations. Consequently, fewer regions become active simultaneously, allowing more aggressive coarsening. In contrast, larger time steps induce larger per-step deformations, which activate more degrees of freedom and reduce coarsening opportunities.
Nevertheless, even with relatively large time steps, our method is able to recover dynamics that closely match the full-resolution simulation while achieving a $2\times$ reduction in total computation time. Unless otherwise stated, we use $\Delta t = 0.01$, a commonly adopted time step size for IPC-based simulations.

\begin{figure}[htbp]
	\centering
	\includegraphics[width=1.0\columnwidth]{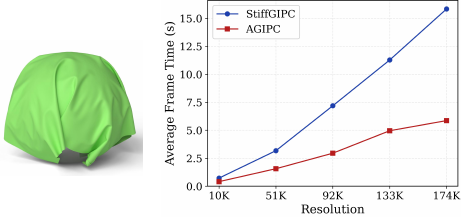}
	\caption{
		\textbf{Scalability test.}
		A cloth with resolutions ranging from 10K to 174K vertices is dropped onto an ABD sphere. As resolution increases, the total runtime of the full-resolution simulation grows superlinearly, while our adaptive method exhibits a much milder increase.
	}
	\label{fig:resolution}
\end{figure}

\paragraph{Scalability Tests.}
We evaluate the scalability of our method by dropping a square cloth with varying spatial resolutions onto an ABD sphere. As shown in \autoref{fig:resolution}, the runtime of the full-resolution simulation increases superlinearly as the mesh resolution grows, whereas our method exhibits substantially improved scaling behavior. While high-resolution meshes are required to capture fine visual details, such detail is not uniformly necessary across all time steps or spatial regions. Our adaptive coarsening strategy exploits this observation by selectively reducing inactive degrees of freedom in both space and time, thereby avoiding unnecessary computation. As a result, our method remains efficient even at high resolutions, while producing results that closely match the full-resolution simulation. 

\begin{table*}[]
	\caption{\textbf{Performance summary compared to StiffGIPC.} We report the number of vertices (\#V), Young’s modulus $E$ (Pa), time step size $\Delta t$, and the simulation framework. Unless otherwise stated, both StiffGIPC and AGIPC use the MAS preconditioner. \texttt{buildGH} denotes the time spent assembling energy gradients and Hessians; \texttt{solve} the time spent solving the linear system; \texttt{CCD} the time for continuous collision detection; \texttt{Linesearch} the time for line search; and \texttt{misc} the time for remaining operations. \texttt{TotalTime} is the overall simulation time, \texttt{TotalNewtonIter} and \texttt{TotalPCGIter} report the total numbers of Newton and PCG iterations, respectively, and \texttt{Speedup} is measured relative to StiffGIPC. All timings are reported in seconds.}
	\label{tab:timecost}
	\resizebox{\textwidth}{!}{%
	\begin{tabular}{c|c|c|c|ccccccccc|c}
	\hline
													  & \#V                   & $E$                       & $\Delta t$             & framework & \texttt{buildGH} & \texttt{solve} & \texttt{CCD} & \texttt{Linesearch} & \texttt{misc} & \texttt{TotalTime} & \texttt{TotalNewtonIter} & \texttt{TotalPCGIter} & \texttt{Speedup}      \\ \hline
	\multirow{2}{*}{\autoref{fig:teaser}}             & \multirow{2}{*}{73K}  & \multirow{2}{*}{3e5}      & \multirow{2}{*}{0.02}  & StiffGIPC & 70.3458          & 346.826        & 22.4561      & 57.0368             & 0.44603       & 497.111            & 8799                     & 1.49622e6             & \multirow{2}{*}{2.04} \\
													  &                       &                           &                        & AGIPC     & 48.1698          & 128.594        & 22.5496      & 43.6347             & 0.262502      & 243.211            & 8705                     & 974688                &                       \\ \hline
	\multirow{10}{*}{\autoref{fig:resolution}}        & \multirow{2}{*}{10K}  & \multirow{10}{*}{1e6}     & \multirow{10}{*}{0.01} & StiffGIPC & 4.92194          & 30.7968        & 1.87036      & 5.69572             & 0.0603376     & 43.3452            & 2314                     & 347425                & \multirow{2}{*}{1.77} \\
													  &                       &                           &                        & AGIPC     & 3.73799          & 15.1909        & 1.86343      & 3.61885             & 0.0578801     & 24.4691            & 2341                     & 249278                &                       \\ \cline{2-2} \cline{5-14} 
													  & \multirow{2}{*}{51K}  &                           &                        & StiffGIPC & 19.4416          & 138.224        & 6.71972      & 25.5912             & 0.105483      & 190.082            & 3374                     & 912276                & \multirow{2}{*}{2.03} \\
													  &                       &                           &                        & AGIPC     & 14.721           & 56.5586        & 6.36533      & 15.7259             & 0.111017      & 93.4818            & 3230                     & 633507                &                       \\ \cline{2-2} \cline{5-14} 
													  & \multirow{2}{*}{92K}  &                           &                        & StiffGIPC & 35.1128          & 335.351        & 12.3552      & 47.5626             & 0.193816      & 430.575            & 3764                     & 1.27198e6             & \multirow{2}{*}{2.44} \\
													  &                       &                           &                        & AGIPC     & 24.062           & 117.066        & 10.6401      & 24.5863             & 0.180908      & 176.535            & 3285                     & 896312                &                       \\ \cline{2-2} \cline{5-14} 
													  & \multirow{2}{*}{133K} &                           &                        & StiffGIPC & 52.1652          & 527.974        & 18.9864      & 76.1894             & 0.326428      & 675.641            & 4035                     & 1.53142e6             & \multirow{2}{*}{2.28} \\
													  &                       &                           &                        & AGIPC     & 41.5033          & 193.656        & 18.3054      & 42.9178             & 0.309966      & 296.692            & 3960                     & 1.2017e6              &                       \\ \cline{2-2} \cline{5-14} 
													  & \multirow{2}{*}{174K} &                           &                        & StiffGIPC & 66.2277          & 755.212        & 24.574       & 102.596             & 0.457173      & 949.067            & 4093                     & 1.74645e6             & \multirow{2}{*}{2.70} \\
													  &                       &                           &                        & AGIPC     & 48.7297          & 250.198        & 22.3166      & 29.3078             & 0.375787      & 350.928            & 3726                     & 1.31649e6             &                       \\ \hline
	\multirow{8}{*}{\autoref{fig:stiffness}}          & \multirow{8}{*}{62K}  & \multirow{2}{*}{1e4}      & \multirow{8}{*}{0.01}  & StiffGIPC & 88.2218          & 419.098        & 35.0655      & 152.657             & 0.544143      & 695.586            & 15300                    & 2.25857e6             & \multirow{2}{*}{1.87} \\
													  &                       &                           &                        & AGIPC     & 62.4948          & 193.076        & 33.2666      & 83.2084             & 0.873126      & 372.919            & 15062                    & 1.51614e6             &                       \\ \cline{3-3} \cline{5-14} 
													  &                       & \multirow{2}{*}{1e5}      &                        & StiffGIPC & 34.8089          & 463.166        & 16.1591      & 45.8648             & 0.241935      & 560.241            & 8593                     & 3.26916e6             & \multirow{2}{*}{2.40} \\
													  &                       &                           &                        & AGIPC     & 25.6385          & 161.442        & 15.8838      & 30.1235             & 0.250318      & 233.338            & 8516                     & 2.07905e6             &                       \\ \cline{3-3} \cline{5-14} 
													  &                       & \multirow{2}{*}{1e6}      &                        & StiffGIPC & 10.3695          & 328.878        & 4.96263      & 12.5261             & 0.0765568     & 356.813            & 2809                     & 2.58436e6             & \multirow{2}{*}{3.01} \\
													  &                       &                           &                        & AGIPC     & 7.8771           & 97.2404        & 5.12317      & 8.27524             & 0.0731677     & 118.589            & 2807                     & 1.47834e6             &                       \\ \cline{3-3} \cline{5-14} 
													  &                       & \multirow{2}{*}{1e7}      &                        & StiffGIPC & 7.888            & 469.561        & 3.52632      & 8.91613             & 0.0560975     & 489.948            & 2151                     & 3.82769e6             & \multirow{2}{*}{3.30} \\
													  &                       &                           &                        & AGIPC     & 7.66572          & 128.361        & 4.64314      & 7.67614             & 0.0714022     & 148.417            & 2736                     & 2.25764e6             &                       \\ \hline
	\multirow{8}{*}{\autoref{fig:dt}}                 & \multirow{8}{*}{50K}  & \multirow{8}{*}{3e5}      & \multirow{2}{*}{0.005} & StiffGIPC & 31.9421          & 279.78         & 11.6789      & 73.7106             & 0.237763      & 397.349            & 8320                     & 1.937e6               & \multirow{2}{*}{2.43} \\
													  &                       &                           &                        & AGIPC     & 22.7476          & 88.7599        & 10.5576      & 41.3489             & 0.195461      & 163.609            & 7718                     & 736588                &                       \\ \cline{4-14} 
													  &                       &                           & \multirow{2}{*}{0.01}  & StiffGIPC & 5.21224          & 70.999         & 2.07459      & 7.41712             & 0.037672      & 85.7406            & 1355                     & 546279                & \multirow{2}{*}{2.07} \\
													  &                       &                           &                        & AGIPC     & 4.28511          & 29.7555        & 2.17194      & 5.25463             & 0.0383109     & 41.5055            & 1439                     & 310713                &                       \\ \cline{4-14} 
													  &                       &                           & \multirow{2}{*}{0.02}  & StiffGIPC & 2.24078          & 52.138         & 0.986562     & 2.48832             & 0.0200349     & 57.8737            & 569                      & 416531                & \multirow{2}{*}{2.16} \\
													  &                       &                           &                        & AGIPC     & 1.85786          & 22.0686        & 0.99883      & 1.79576             & 0.0175809     & 26.7386            & 622                      & 252051                &                       \\ \cline{4-14} 
													  &                       &                           & \multirow{2}{*}{0.04}  & StiffGIPC & 1.38184          & 52.6121        & 0.648034     & 1.48074             & 0.0125185     & 56.1352            & 348                      & 441139                & \multirow{2}{*}{2.07} \\
													  &                       &                           &                        & AGIPC     & 1.13823          & 24.3456        & 0.643354     & 1.0402              & 0.0108994     & 27.1783            & 381                      & 286114                &                       \\ \hline
	\multirow{2}{*}{\autoref{fig:balls_and_xmas} (a)} & \multirow{2}{*}{185K} & \multirow{2}{*}{mixed}    & \multirow{2}{*}{0.01}  & StiffGIPC & 49.3885          & 2663.14        & 41.519       & 67.2764             & 0.175731      & 2821.50            & 5313                     & 7.86873e6             & \multirow{2}{*}{3.15} \\
													  &                       &                           &                        & AGIPC     & 50.6319          & 733.995        & 54.1771      & 55.8605             & 0.223638      & 894.888            & 7617                     & 7.22941e6             &                       \\ \hline
	\multirow{2}{*}{\autoref{fig:balls_and_xmas} (b)} & \multirow{2}{*}{127K} & \multirow{2}{*}{mixed}    & \multirow{2}{*}{0.01}  & StiffGIPC & 28.0893          & 632.316        & 11.2302      & 22.7132             & 0.120316      & 694.469            & 3944                     & 2.45136e6             & \multirow{2}{*}{2.24} \\
													  &                       &                           &                        & AGIPC     & 20.6969          & 263.428        & 12.1944      & 13.7154             & 0.110192      & 310.145            & 3989                     & 3.63167e6             &                       \\ \hline
	\multirow{2}{*}{\autoref{fig:dolphin}}            & \multirow{2}{*}{8K}   & \multirow{2}{*}{1e4}      & \multirow{2}{*}{0.01}  & StiffGIPC & 29.1452          & 140.1          & 28.6722      & 82.1661             & 0.334575      & 280.418            & 13789                    & 1.72152e6             & \multirow{2}{*}{1.41} \\
													  &                       &                           &                        & AGIPC     & 25.722           & 81.2531        & 37.0044      & 54.4989             & 0.257074      & 198.735            & 14933                    & 2.57214e6             &                       \\ \hline
	\multirow{2}{*}{\autoref{fig:twist}}              & \multirow{2}{*}{45K}  & \multirow{2}{*}{1e7}      & \multirow{2}{*}{0.01}  & StiffGIPC & 611.813          & 9076.36        & 229.418      & 362.195             & 4.70331       & 10284.5            & 71125                    & 1.15179e8             & \multirow{2}{*}{2.87} \\
													  &                       &                           &                        & AGIPC     & 319.219          & 2954.11        & 166.393      & 134.259             & 5.17297       & 3579.15            & 58032                    & 6.05664e7             &                       \\ \hline
	\multirow{2}{*}{\autoref{fig:sig_thank}}          & \multirow{2}{*}{172K} & \multirow{2}{*}{3e5}      & \multirow{2}{*}{0.02}  & StiffGIPC & 342.498          & 929.733        & 61.4486      & 379.301             & 3.37835       & 1716.36            & 6476                     & 2.20141e6             & \multirow{2}{*}{2.00} \\
													  &                       &                           &                        & AGIPC     & 248.611          & 313.131        & 57.4081      & 236.323             & 3.54204       & 859.015            & 7143                     & 5.79561e6             &                       \\ \hline
	\end{tabular}%
	}
\end{table*}

	\begin{figure}[htbp]
	\centering
	\includegraphics[width=\columnwidth]{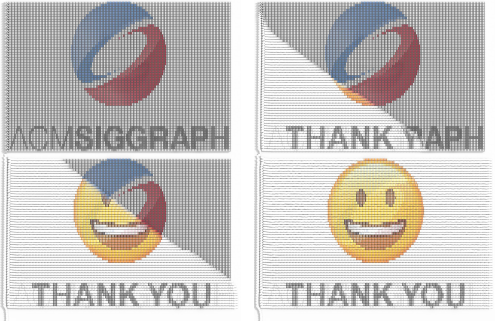}
	\caption{
		\textbf{Domino SIGGRAPH THANK YOU scene.} A large domino scene with over 9K deformable cards and 172K vertices. As the cards fall sequentially, the pattern transitions from SIGGRAPH to THANK YOU.
	}
	\label{fig:sig_thank}
	\end{figure}

\paragraph{Overall Performance.}
\autoref{tab:timecost} summarizes the detailed runtime breakdown of our AGIPC framework in comparison with StiffGIPC. 
In addition to introducing the algebraic adaptive coarsening framework, we further optimize
some stages of the pipeline other than the linear solver such as the Hessian assembly and collision detection. 
First, we accelerate the Hessian assembly by exploiting matrix symmetry: both FEM elasticity and IPC contact/friction Hessians are symmetric, so we store and accumulate only the diagonal and upper‑triangular entries during assembly. This reduces memory traffic and assembly cost for both the fine‑mesh Hessian and the subsequent coarse system. Second, we optimize collision detection using a stack‑less BVH traversal, in contrast to the stack‑based approach in StiffGIPC. Combined with all these optimizations, AGIPC achieves up to $3.3\times$ speedup over StiffGIPC while maintaining visually identical results in all experiments. The detailed coarsening rate of our framework is provided in the supplementary document.

    \section{Conclusion and Discussion}
In this paper, we introduced algebraic adaptive in‑solve coarsening, a GPU‑friendly method that accelerates implicitly time-stepped IPC simulation by dynamically coarsening the linear system within each Newton iteration. Our method preserves IPC’s barrier continuity by avoiding explicit topological changes, while enabling efficient parallel reductions that directly assemble a coarser system from fine mesh data. Guided by a Green strain increment criterion, edges are adaptively collapsed to drastically reduce DoF, while selective affine embedding maintains rotational motion. Experiments show that our framework delivers up to $3\times$ speedup over state‑of‑the‑art GPU IPC framework, with visually identical results. The method is robust across a range of stiffnesses, deformations, and contact scenarios. We believe this work offers a practical GPU‑friendly foundation for adaptive simulation, with potential applications in graphics, engineering, and scientific computing.

\paragraph{Limitations and future work.}
Our method currently employs a fixed Green strain increment threshold for adaptive coarsening in most experiments. While this choice works well across a wide range of scenarios, it is not optimal for all deformation regimes and can be sensitive to numerical noise in certain cases (see \autoref{fig:twist}). A promising direction for future work is to adapt this threshold dynamically, for example based on local geometric quality, solver state, or learned signals from the simulation history, which could further improve robustness and efficiency.

Second, although our algebraic coarsening approach reduces the number of degrees of freedom in the linear system, it does not modify the underlying mesh discretization. As a result, Hessian assembly and collision detection still have to be performed on the full-resolution mesh. In some scenarios, such as \autoref{fig:dolphin} and \autoref{fig:sig_thank} reported in \autoref{tab:timecost}, these stages dominate the overall runtime, even with our optimized framework, diminishing the relative advantage of adaptive coarsening. It will be meaningful to address this open problem with more aggressive GPU-oriented strategies for reduction, collision detection, and contact handling.

Finally, our current adaptive coarsening strategy is driven primarily by kinematic coherence and relies on topological connectivity within individual objects. This limits aggregation across object boundaries. In highly collision-intensive scenarios involving many small objects, such as the domino scene in \autoref{fig:sig_thank}, coarsening is confined within each object. Extending the aggregation strategy to incorporate collision-induced connectivity, for example by treating persistent contact pairs as candidates for aggregation, could enable more global coarsening and further improve performance in these challenging settings.

\begin{acks}
We sincerely thank reviewers for their insightful suggestions. Taku Komura acknowledges partial support from the Innovation and Technology Commission of the HKSAR Government under the ITSP-Platform grants (Refs. ITS/335/23FP and ITS/469/24FP). Minchen Li acknowledges partial support from a Junior Faculty Startup Fund from Carnegie Mellon University and gift funding from Genesis AI. 
\end{acks}

	\bibliographystyle{ACM-Reference-Format}
	\bibliography{ref}



\end{document}


\title{AGIPC: Adaptive In-Solve Algebraic Coarsening for GPU IPC \\Supplemental Document}
    
        \author{Xuan Wang}
    \orcid{0009-0004-7061-8321}
	\email{xuan-wang@connect.hku.hk}
	\affiliation{%
		\institution{The University of Hong Kong}
		\country{Hong Kong SAR}
	}

    \author{Zhaofeng Luo}
    \orcid{0009-0007-8947-9108}
	\email{zhaofen2@andrew.cmu.edu}
	\affiliation{%
		\institution{Carnegie Mellon University}
		\country{USA}
	}

    \author{Minchen Li}
    \orcid{0000-0001-9868-7311}
	\email{minchernl@gmail.com}
	\affiliation{%
		\institution{Carnegie Mellon University, Genesis AI}
		\country{USA}
	}
        
	\author{Taku Komura}
	\orcid{0000-0002-2729-5860}
	\email{taku@cs.hku.hk}
	\affiliation{%
		\institution{The University of Hong Kong}
		\country{Hong Kong SAR}
	}

    \author{Kemeng Huang}
    \authornote{Corresponding author and project lead}
	\orcid{0000-0001-9147-2289}
	\email{kmhuang@connect.hku.hk}
	\email{kmhuang819@gmail.com}
	\affiliation{%
		\institution{The University of Hong Kong}
		\country{Hong Kong SAR}
	}
        
	\renewcommand{\shortauthors}{Xuan Wang, Zhaofeng Luo, Minchen Li, Taku Komura, Kemeng Huang}

	\begin{abstract}
		{This document provides supplementary details about the technical background, derivations, implementation, and experiments of our methods.}
	\end{abstract}

	\ccsdesc[500]{Computing methodologies~Physical simulation}
	\ccsdesc[300]{Computing methodologies~Parallel algorithms}
	
	\keywords{GPU IPC, Adaptive Coarsening, Green Strain, Affine Embedding}
    \maketitle
    \tableofcontents

\section{Fine‑to‑Coarse Mapping via Parallel Hashing}
\label{app:hash_mapping}

We construct the fine‑to‑coarse mapping using a parallel hashing scheme inspired by GPU MAS \cite{stiffgipc}. The mesh nodes are first partitioned into contiguous groups, each processed by a CUDA Warp (a Warp may handle multiple groups). Within a group, every node builds a compact bit‑hash encoding its connectivity to collapsible neighbors. Connected components are identified by iteratively merging these hashes via bitwise‑OR across neighbors. The position of the first set bit in the final hash determines the node’s local coarse index within the group; a global prefix‑sum over the number of coarse super‑nodes per group then yields unique global coarse indices. The full procedure is detailed in Algorithms \ref{alg:L0_Construction} and \ref{alg:expand_connection}.

\begin{algorithm}
		\caption{Direct neighbor hash encoding}\label{alg:L0_Construction}
		\begin{algorithmic}[1]
			\vspace{0.1cm}
			\For{$node\_id = 0, 1, ..., node\_number-1$} \textbf{in parallel}
			\State $group\_id = node\_id/group\_size$
			\State $lane\_id = node\_id\%group\_size$
			\State $con\_hashs[node\_id] = 1<<lane\_id$ 
			\For{$neighbor\_id \leftarrow \text{remained adjacent neighbor index}$}
            \State $t \leftarrow \text{get corresponding edge tag}$
            \If{$t == 0$}
			\State $continue$
			\EndIf
			\State $neighbor\_group\_id = neighbor\_id/group\_size$
			\If{neighbor\_group\_id == group\_id}

			\State $neighbor\_lane\_id = neighbor\_id\%group\_size$
			\State $con\_hashs[node\_id]\, \text{|=}\, (1<<neighbor\_lane\_id)$

			\EndIf
			\EndFor
			\State update remained neighbor info for node $node\_id$
			\EndFor
		\end{algorithmic}
	\end{algorithm}

\begin{figure}[htbp]
\centering
\begin{subfigure}[b]{0.46\textwidth}
\includegraphics[width=\columnwidth, trim=0 0 0 0, clip]{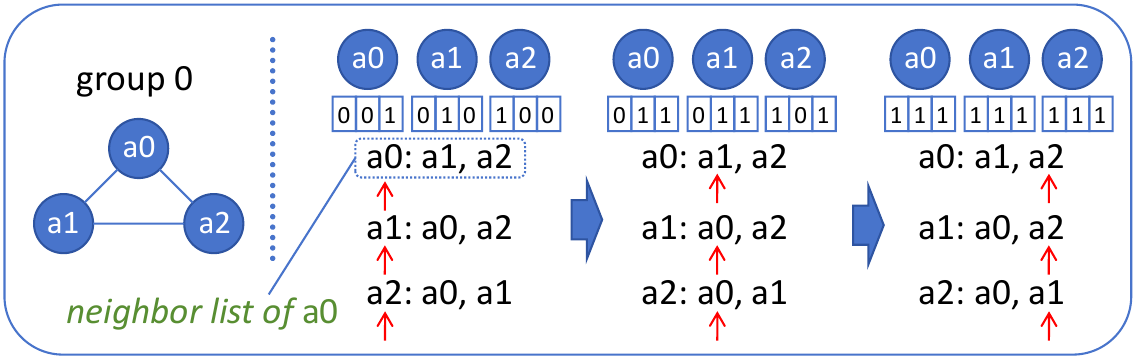}
\end{subfigure}
\caption{
\textbf{A direct neighbor hash encoding example.} We illustrate the procedure of Algorithm~\ref{alg:L0_Construction} using group~0 from Figure~2 of main paper. First, each node's hash is initialized by setting the bit corresponding to its local position within the group. We then traverse the neighbors of each target node and, for every neighbor belonging to the same group, set the bit associated with that neighbor's local index.
}
\label{fig:neighbor_travesal}
\end{figure}

\subsection{Adaptive Coarsening via Edge Tags}

Adaptivity is controlled by a binary tag $\tau_e \in \{0,1\}$ assigned to each edge $e$. By default, $\tau_e = 1$, marking the edge as collapsible. When a local deformation criterion is met, the tag is set to $\tau_e = 0$, protecting the edge and preserving local detail. This tag‑based design avoids explicit coarse representations, common in traditional remeshing or subspace methods, which would require maintaining dynamic changed data structure, leading to irregular memory access and necessitating conservative, oversized memory pre‑allocation on GPU to guard against runtime reallocation during dynamic updates. Instead, we treat coarsening as a stateless, per‑step process that starts each Newton iteration from the original fine mesh. The static underlying topology of fine mesh ensures predictable, contiguous memory access, which is crucial for high‑performance GPU execution.

\begin{algorithm}
		\caption{Indirect neighbor hashing integration and mapping}\label{alg:expand_connection}
		\begin{algorithmic}[1]
			\vspace{0.1cm}
			\For{$node\_id = 0, 1, ..., node\_number-1$} \textbf{in parallel}
			\State $group\_id = node\_id/group\_size$
                \State $b\_tid = node\_id\%blockSize$
			\State $b\_group\_id = (b\_tid)/group\_size$
			\State $lane\_id = node\_id\%group\_size$
			\State $Hashs[blockSize] \gets$ \textbf{shared memory}
			\State $counts[blockSize/group\_size] \gets$ \textbf{shared memory}
            \State $electHash[blockSize/group\_size] \gets$ \textbf{shared memory}
            \If{$lane\_id == 0$} 
			\State $counts[b\_group\_id] = electHash[b\_group\_id]= 0$
			\EndIf
			\State $conHash = con\_hashs[node\_id]$
			\State $Hashs[b\_tid] = conHash$
			\State $visited = (1<<lane\_id)$
			\While{$conHash \neq (1<<group\_size)-1$}
			\State $todo = (visited \oplus conHash)$
                \State \textbf{If} $todo == 0$ \textbf{then} $break$
			\State $next = \textbf{\_\_ffs}(todo)-1$ 
			\State $visited\, \text{|=}\, (1<<next)$
			\State $conHash\, \text{|=}\, Hashs[next+b\_group\_id\times group\_size]$
			\EndWhile
			
			\State $con\_hashs[node\_id] = conHash$
			
			\State $prefix = \textbf{\_\_popc}(conHash\, \textbf{And}\, ((1<<lane\_id)-1))$
			\Statex \hspace{0.4cm} {\color[rgb]{0,0.61,0.33}{// count the coarse nodes in each group:}}
			\If{$prefix == 0$}
			\State $\textbf{atomicAdd}(\&counts[b\_group\_id],1)$
            \State $\textbf{atomicOr }(\&electHash[b\_group\_id],1<<lane\_id)$
			\EndIf
            \State $mask = electHash[b\_group\_id]\, \&\, ((1<<lane\_id)-1)$
			\State $P[node\_id] = \textbf{\_\_popc}(mask)$
			\If{$lane\_id == 0$}
			\State $coarse\_node\_num[group\_id] = counts[b\_group\_id]$
			\EndIf
			
			\EndFor

            \State $O \gets\,$ExclusiveSum($coarse\_node\_num$)
            \For{$node\_id = 0, 1, ..., node\_number-1$} \textbf{in parallel}
            \State $group\_id = node\_id/group\_size$
            \State $map[node\_id] = O[group\_id]+P[node\_id]$
            \EndFor
		\end{algorithmic}
	\end{algorithm}

\subsection{Illustrative Example of Hash Construction}

To illustrate the hashing process, we refer to the example in Figure 2 of the main paper. Assuming a warp size of 3 and a group size of 3, nodes are grouped sequentially by their index. Each node constructs a bit‑hash based on its relative position within its group as shown in \autoref{fig:neighbor_travesal}. We then traverse each node’s neighbors (lines 5–15, Algorithm~\ref{alg:L0_Construction}). If a neighbor resides in the same group and the connecting edge is collapsible, the bit corresponding to that neighbor’s intra‑group index is set to 1 (lines 12–15). In this example, since nodes $\{a0,a1,a2\}$ are mutually connected, their hashes all become `111`.

\begin{figure*}[htbp]
\centering
\begin{subfigure}[b]{0.95\textwidth}
\includegraphics[width=\columnwidth, trim=0 0 0 0, clip]{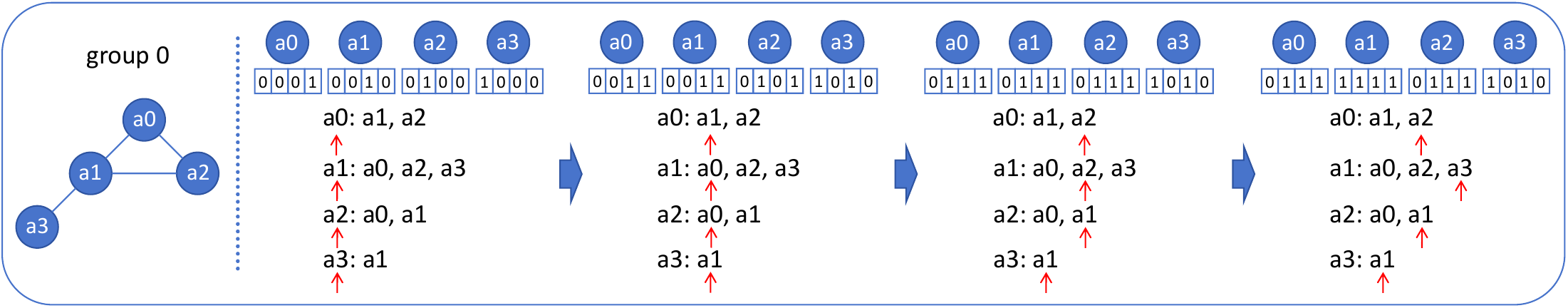}
\end{subfigure}
\caption{
\textbf{A hash encoding example under non‑mutual connectivity.} We illustrate the same construction procedure of Algorithm~\ref{alg:L0_Construction} using an example where connectivity is not mutual.
}
\label{fig:neighbor_travesal2}
\end{figure*}

More generally, nodes are often linked indirectly. For instance, in a group of size 4 as shown in \autoref{fig:neighbor_travesal2}), Algorithm~\ref{alg:L0_Construction} produces the hashes $\{0111, 1111, 0111, 1010\}$. Even though ${a0, a1, a2, a3}$ are not directly mutually connected, they form a fully connected component. To capture this, we propagate connectivity information by iteratively merging neighbor hashes with bitwise‑OR (lines 12–22, Algorithm~\ref{alg:expand_connection}). For $a0$, bit 1 is set, so we combine its hash with $a1$’s hash `1111`, yielding `1111`. After propagating across the group, all nodes converge to the same hash `1111`, indicating they form a connected component. All nodes with identical final hashes are merged into a single coarse super‑node. 

Since a single 32‑bit interger can represent at most 32 neighbors, matching the thread count of a CUDA Warp, groups larger than 32 cannot be processed in a single step. We therefore apply the same encoding scheme in Algorithms \autoref{alg:L0_Construction} and \autoref{alg:expand_connection} recursively: each level of coarse nodes becomes the input to the next level. This hierarchical process continues until a minimal set of DoF is obtained, as illustrated in the right side of Figure 2 in the main paper, where coarse nodes $b0$, $b1$, and $b2$ are merged into a single node $C0$.

\subsection{Fine‑to‑Coarse Mapping and adaptive coarsening}
The previous examples illustrate the default coarsening process, in which all edges are tagged 1 (collapsible). During simulation, certain regions must retain fine scale detail, requiring that some edges be protected from collapse by setting their tags to 0. When an edge tag is set to 0, the corresponding hash bit remains unset (lines 8–10, Algorithm~\ref{alg:L0_Construction}). This breaks the local connectivity, resulting in distinct hash values and a correspondingly finer coarse mesh. 

Consider the example in Figure~3 of the main paper, where all edges adjacent to nodes ${a3, a4, a5}$ are tagged 0, and the group size is 3. Their resulting hashes are $\{001, 010, 100\}$. Because no neighbor bits are set, the hashes remain orthogonal, and each node is preserved individually in the coarse mesh. The coarse index of a node is determined by the position of the first ‘1’ in its hash (lines 23, 26, 28, Algorithm~\ref{alg:expand_connection}). For instance, $a3$ with hash '001' has its first set bit at position 0, giving it order 0 within the group; $a4$ ('010') becomes order 1, and $a5$ ('100') becomes order 2.

The global coarse index is obtained by adding a prefix‑sum over the number of coarse nodes in each group (lines 29 and 34–38, Algorithm~\ref{alg:expand_connection}). Since the first group contains one coarse node \footnote{Importantly, the number of coarse nodes in each group equals the number of distinct hash values within that group (lines 23 and 25, Algorithm~\ref{alg:expand_connection}).}, the final global indices for $\{a3, a4, a5\}$ become $\{1, 2, 3\}$. We store this relationship in a mapping array that links each fine node to its corresponding coarse node (line 37, Algorithm~\ref{alg:expand_connection}). This mapping array is subsequently used to construct the coarse linear system and to prolongate the coarse solution back to the fine mesh.

In the default case shown in Figure~2 of the main paper, where all edges are collapsible, the mapping reduces to the constant function $Map(i) = 0$ for all fine nodes $i$. To ensure that nodes within each group are well connected and can be aggregated efficiently, a METIS based sorting method \cite{stiffgipc} is applied. This process enables the fine mesh to be coarsened to a minimal set of degrees of freedom, which in the limit becomes a single coarse node.

\section{Coarse Hessian Assembly on GPU}
For clarity and self‑containment, we briefly review the fast hash‑based reduction method used in StiffGIPC \cite{stiffgipc}. In BCOO (Blocked Coordinate) format, each non‑zero entry of the Hessian is stored as a triplet $(i, j, \mathbf{B}_{ij})$, where $i$ and $j$ are global row and column indices and $\mathbf{B}_{ij}$ is the corresponding $3\times 3$ block matrix. To merge duplicate entries efficiently, each triplet is assigned a 64‑bit hash key computed as $(i \ll 32) \mid j$. Triplets with the same $(i,j)$ indices produce identical hash keys and thus correspond to the same matrix location.

To assemble the global Hessian, the array of triplets is first sorted by their hash keys, grouping together all entries that map to the same $(i,j)$ position. Within each group of identical keys, the $3\times 3$ blocks are summed via a parallel segmented reduction, implemented using CUDA Warp level primitives for high GPU efficiency. The result is a unique set of triplets in BCOO format, with accumulated block contributions ready for the linear solver.

Our coarse linear system is constructed from the fine‑mesh Hessian already assembled via the fast hash‑based reduction. Starting from the unique BCOO triplets of the fine system, each triplet $(i, j, \mathbf{B}_{ij})$ is mapped to coarse indices using the fine‑to‑coarse mapping function: $(map(i), map(j), \mathbf{B}_{ij})$. Because multiple fine nodes may map to the same coarse node, duplicate entries can reappear after this mapping. We therefore perform a second hash‑based reduction, now using the hash key $(map(i) \ll 32) \mid map(j)$. The resulting unique set of triplets constitutes the coarse Hessian in BCOO format, ready for the linear solver.

\begin{algorithm}
\caption{Deciding DoF for coarse nodes}
\label{alg:affine_mapping1}
\begin{algorithmic}[1]
\vspace{0.1cm}
\For{$i = 0, 1, \dots, coarse\_node\_num - 1$} \textbf{in parallel}
\If{the number of fine nodes mapped to $i$ exceeds 32}
\State $coarse\_dof[i] = 12$ \Comment{12 DoF with affine mapping}
\State $reorder\_hash\_key[i] = i + coarse\_node\_num$ 
\Else
\State $coarse\_dof[i] = 3$ \Comment{3 DoF with identity mapping}
\State $reorder\_hash\_key[i] = i$
\EndIf
\EndFor
\Statex{\color[rgb]{0,0.61,0.33}{// Reorder nodes so that 3DoF nodes precede 12DoF nodes}}
\State SortPairs($reorder\_hash\_key$, $indices$)
\For{$i = 0, 1, \dots, coarse\_node\_num - 1$} \textbf{in parallel}
\State $new\_map[i] = map[indices[i]]$
\EndFor
\end{algorithmic}
\end{algorithm}

\begin{algorithm}
\caption{Transforming fine Hessian triplets}
\label{alg:affine_mapping2}
\begin{algorithmic}[1]
\vspace{0.1cm}
\Statex{\color[rgb]{0,0.61,0.33}{// Determine coarse block type and block multiplicity}}
\For{each triplet $T_{th} = (i, j, \mathbf{B}_{ij})$} \textbf{in parallel}
\If{nodes $new\_map[i]$ and $new\_map[j]$ are 3DoF}
\State $\mathbf{H}^{3\times3}_{ij} = \mathbf{B}_{ij}$ \Comment{$3\times3$}
\State $block\_counts[T_{th}] = 1$
\ElsIf{nodes $new\_map[i]$ and $new\_map[j]$ are 12DoF}
\State $\mathbf{H}^{12\times12}_{ij} = \mathbf{A}_i \mathbf{B}_{ij} \mathbf{A}_j^T$ \Comment{$12\times12$}
\State $block\_counts[T_{th}] = 16$
\ElsIf{$new\_map[r]$ is 3DoF and $new\_map[c]$ is 12DoF}
\State $\mathbf{H}^{3\times12}_{ij} = \mathbf{B}_{ij} \mathbf{A}_j^T$ \Comment{$3\times12$}
\State $block\_counts[T_{th}] = 4$
\ElsIf{$new\_map[r]$ is 12DoF and $new\_map[c]$ is 3DoF}
\State $\mathbf{H}^{12\times3}_{ij} = \mathbf{A}_i \mathbf{B}_{ij}$ \Comment{$12\times3$}
\State $block\_counts[T_{th}] = 4$
\EndIf
\EndFor
\Statex{\color[rgb]{0,0.61,0.33}{// Compute offsets for flattened $3\times3$ blocks}}
\State ExclusiveSum($block\_counts$, $block\_offsets$)

\For{each $T_{th} = (i, j, \mathbf{H}_{ij})$} \textbf{in parallel}
\For{each $3\times3$ block $\mathbf{h}^{3\times3} \text{ in } \mathbf{H}_{ij}$}
\State $K_{th}\leftarrow$row-major-order block index in $\mathbf{H}_{ij}$ 
\State $global\_triplet\_{id} = block\_offsets[T_{th}]+K_{th}$
\State $r\leftarrow\text{calculate global row index in } \mathbf{H}_c$
\State $c\leftarrow\text{calculate global column index in } \mathbf{H}_c$
\State store the Triplet $\{r, c, \mathbf{h}^{3\times3}\}$ using $global\_triplet\_{id}$
\EndFor

\EndFor
\end{algorithmic}
\end{algorithm}

\section{Modified Affine Approximation Mapping}
\label{app:affine_mapping}

Extending the two‑level hash‑based reduction method of \citet{stiffgipc} to support affine‑enriched coarse nodes is not straightforward, as the presence of heterogeneous block sizes prevents the original uniform reduction scheme from being applied directly. In the fine‑level Hessian, all entries $\mathbf{B}_{ij}$ are uniform $3\times3$ blocks, but coarse nodes with affine approximations possess 12 DoF, leading to four possible coarse block types: $3\times3$ (both nodes with 3 DoF), $3\times12$ or $12\times3$ (mixed DoF), and $12\times12$ (both nodes with 12 DoF). To maintain efficiency in parallel hashing and reduction, we decompose mixed size blocks into collections of standard $3\times3$ blocks, following a strategy analogous to ABD–FEM coupling \cite{stiffgipc}.


As summarized in Algorithm~\ref{alg:affine_mapping1}, we first classify each coarse node as either identity-mapped (3 DoF) or affine-mapped (12 DoF), based on the number of fine nodes it aggregates. In practice, we assign affine mappings to coarse nodes whose aggregate size exceeds a predefined threshold (lines 1-9). After classification, we reorder the coarse nodes such that all 3-DoF nodes precede the 12-DoF nodes (lines 10-13). This reordering induces a block-structured coarse Hessian of the form
\begin{equation}
\mathbf{H}_c =
\begin{bmatrix}
\mathbf{H}_{3dof} & \mathbf{H}_{mixed} \\
\mathbf{H}_{mixed} & \mathbf{H}_{12dof}
\end{bmatrix},
\end{equation}
which simplifies subsequent indexing and block flattening.

For each fine-level $3 \times 3$ Hessian block $\mathbf{B}_{ij}$, we first apply appropriate transformation based on how its row and column vertices $i$ and $j$ are mapped. Each fine node that is mapped to a 12-DoF coarse node is associated with a $12 \times 3$ affine transformation matrix $\mathbf{A} = \bar{\mathbf{X}} \otimes \mathbf{I}_3$, where $\bar{\mathbf{X}}$ denotes the homogeneous rest-pose coordinate of the fine node. Depending on the mapping combinations of the two vertices, there are four possible types of transformed blocks $\mathbf{H}_{ij}$ with different matrix sizes (lines 1-15, Algorithm~\ref{alg:affine_mapping2}). To flatten these mixed-size blocks into standard $3 \times 3$ blocks, we first count the number of $3 \times 3$ sub-blocks contained in $\mathbf{H}_{ij}$ and perform an exclusive prefix sum to determine its global offset in the triplet array (line 16). Then for each $3 \times 3$ sub-block of $\mathbf{H}_{ij}$, we store the corresponding triplet entries starting at this global offset (lines 17-25). The computation of global indices depends on the DoF of the mapped nodes. Since we have reordered the coarse nodes, placing all 3-DoF nodes before 12-DoF nodes, if $new\_map[i]$ (or $new\_map[j]$) corresponds to a 3-DoF coarse node, its global index is simply $new\_map[i]$ (or $new\_map[j]$). However, if a mapped node corresponds to a 12-DoF coarse node, its global index must be expanded accordingly. Specifically, let $n_{3dof}$ denote the total number of 3-DoF coarse nodes. If $new\_map[i]$ is a 12-DoF node, the global row index is computed as
\begin{equation}
    r = n_{3dof} + \left(new\_map[i] - n_{3dof}\right) \times 4 + \left\lfloor\frac{K_{th}}{4}\right\rfloor.
\end{equation}
Similary, if $new\_map[j]$ is a 12-DoF node, the global column index is 
\begin{equation}
    c = n_{3dof} + \left(new\_map[j] - n_{3dof}\right) \times 4 + (K_{th} \% 4).
\end{equation}
After collecting all the coarse Hessian BCOO triplets, we can remove the duplicate triplets using the fast hash‑based reduction method mentioned in the last section.

\section{Additional Statistics and Experiment Results}

In \autoref{tab:DoF}, we report the DoF statistics for the scenes in the main paper. As shown, our method effectively reduces the number of active DoF efficiently on average per Newton iteration. Notably, a smaller active DoF ratio does not always translate into a proportionally higher runtime speedup. In practice, the performance gain depends on multiple factors, including the underlying scene dynamics, contact complexity, and the distribution of affine versus identity-mapped coarse nodes. Therefore, the active ratio should be interpreted as an indicator of potential efficiency rather than a direct predictor of overall simulation speed.

\begin{table}[htbp]
\caption{\textbf{DoF statistics for the scenes in the main paper.} The number of fine DoF is computed as three times the total number of FEM vertices. For each scene, we report the average number of 3DoF nodes and 12DoF nodes per Newton iteration. The total number of coarse DoF is computed as $\text{(3DoF node)} \times 3 + \text{(12DoF node)} \times 12$. The active ratio is defined as the ratio between the number of coarse DoF and the number of fine DoF. ‘Figure X’ refers to the corresponding figures in the main paper.}
\label{tab:DoF}
\resizebox{0.95\columnwidth}{!}{%
\begin{tabular}{c|c|c|c|c|c}
\hline
									& fine DoF & 3DoF node & 12DoF node & coarse DoF & active ratio \\ \hline
Figure 1             & 218568  & 14675     & 113        & 45381      & 0.21         \\ \hline
Figure 15, 10K    & 31212    & 3517      & 20         & 10791      & 0.35         \\
Figure 15, 51K    & 153228   & 15170     & 83         & 46506      & 0.30         \\
Figure 15, 92K    & 277248   & 29451     & 138        & 90009      & 0.32         \\
Figure 15, 133K   & 399675   & 35502     & 166        & 108498     & 0.27         \\
Figure 15, 174K   & 521667   & 50839     & 240        & 155397     & 0.30         \\ \hline
Figure 13, 1e4     & 185007   & 13207     & 172        & 41685      & 0.23         \\
Figure 13, 1e5     & 185007   & 13953     & 243        & 44775      & 0.24         \\
Figure 13, 1e6     & 185007   & 15177     & 171        & 47583      & 0.26         \\
Figure 13, 1e7     & 185007   & 8473      & 145        & 27159      & 0.15         \\ \hline
Figure 14, 0.005          & 149556   & 5887      & 25         & 17961      & 0.12         \\
Figure 14, 0.01           & 149556   & 15774     & 39         & 47790      & 0.32         \\
Figure 14, 0.02           & 149556   & 23810     & 40         & 71910      & 0.48         \\
Figure 14, 0.04           & 149556   & 33768     & 31         & 101676     & 0.68         \\ \hline
Figure 12 (a) & 555213   & 25848     & 476        & 83256      & 0.15         \\ \hline
Figure 12 (b) & 366837   & 17697     & 235        & 55911      & 0.15         \\ \hline
Figure 9            & 25197    & 5506      & 7          & 16602      & 0.66         \\ \hline
Figure 10              & 135000   & 15471     & 96         & 47565      & 0.35         \\ \hline
Figure 16          & 517152   & 27922     & 0          & 83766      & 0.16         \\ \hline
\end{tabular}%
}
\end{table}

\begin{figure}[htbp]
\centering
\includegraphics[width=\columnwidth]{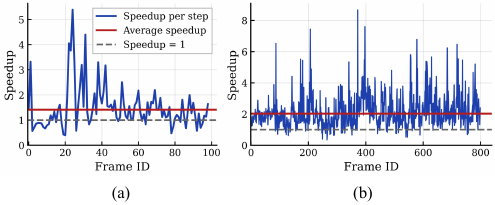}
\caption{\color{black}
\textbf{Per-timestep speedup analysis.}
We plot the per-timestep speedup of our framework relative to StiffGIPC. (a) In the dolphin scenario (Figure 9 in the main paper), peaks represent phases where local deformation allows for more aggressive coarsening. (b) In the teaser scene (Figure 1 in the main paper), the speedup remains more consistent across the simulation duration.
}
\label{fig:per_timestep_speedup}
\end{figure}

{\color{black}
\paragraph{Per-timestep Performance.}
Beyond the overall performance, \autoref{fig:per_timestep_speedup} shows our speedup across individual time steps. In the dolphin scenario (Figure 9 in the main paper), while the overall speedup is moderate, our framework achieves transient speedups exceeding $5\times$ during specific intervals. This variation highlights that our adaptive mechanism is highly responsive to the instantaneous physical complexity of the scene, dynamically adjusting DoF to match the requirement of the current state.
}

\begin{table}[htbp]
	\caption{\color{black} \textbf{Performance in the heterogeneous case.} We report the total numbers of Newton iterations (\texttt{TotalNewtonIter}) and PCG iterations (\texttt{TotalPCGIter}), the total simulation time (\texttt{TotalTime}), and the speedup (\texttt{Speedup}) for the heterogeneous setup in \autoref{fig:heterogeneous_ball}. Both StiffGIPC and AGIPC use the MAS preconditioner.}
	\label{tab:heterogeneous}
	\resizebox{\columnwidth}{!}{%
	\begin{tabular}{c|cc|cc|cc|c}
	\hline
	\multirow{2}{*}{} & \multicolumn{2}{c|}{\texttt{TotalNewtonIter}} & \multicolumn{2}{c|}{\texttt{TotalPCGIter}} & \multicolumn{2}{c|}{\texttt{TotalTime}} & \multirow{2}{*}{\texttt{Speedup}} \\ \cline{2-7}
					  & StiffGIPC               & AGIPC               & StiffGIPC            & AGIPC               & StiffGIPC           & AGIPC             &                                   \\ \hline
	blue              & 2056                    & 2148                & 2.1332e6             & 1.01346e6           & 281.046             & 79.2822           & 3.54                              \\ \hline
	orange            & 1830                    & 2562                & 3.2876e6             & 1.91274e6           & 409.255             & 135.826           & 3.01                              \\ \hline
	green             & 13889                   & 17057               & 2.60274e7            & 1.65155e7           & 3243.42             & 989.322           & 3.28                              \\ \hline
	\end{tabular}%
	}
\end{table}

\begin{figure}[htbp]
\centering
\includegraphics[width=\columnwidth]{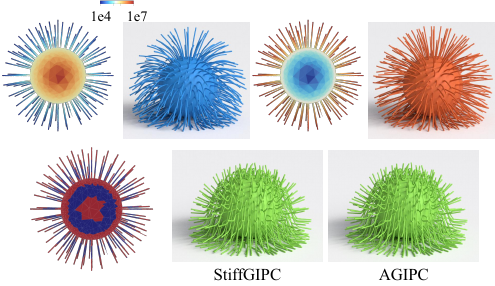}
\caption{\color{black}
\textbf{Heterogeneous materials.}
We evaluate our framework on the squishy ball with spatially varying stiffness. (Top) Two continuous gradients where Young's modulus varies radially. (Bottom) A discrete configuration with interleaved stiffness levels of $10^4$ and $10^7$ Pa in a spherical pattern. Despite the increased numerical complexity of these heterogeneous systems, our framework maintains consistent visual fidelity and computational speedups.
}
\label{fig:heterogeneous_ball}
\end{figure}

{\color{black}\paragraph{Heterogeneous Materials.}
We evaluate the robustness of our framework across different material distributions by modulating the internal stiffness of the squishy ball (see \autoref{fig:heterogeneous_ball} and \autoref{tab:heterogeneous}). In heterogeneous material scenarios, despite exhibiting similar visual dynamics, simulation costs often vary due to differences in the resulting linear systems. We test three configurations: two with continuous stiffness gradients and one with a discrete, high-contrast distribution. In all cases, our method consistently achieves a speedup of over $3\times$, demonstrating that our adaptive framework provides robust acceleration across different material distributions.
}

\begin{figure}[htbp]
\centering
\includegraphics[width=\columnwidth]{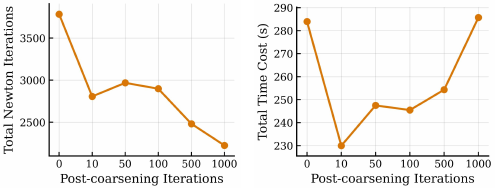}
\caption{\color{black}
\textbf{Impact of post-coarsening iterations on solver performance.}
We sweep the cap on post-coarsening PCG iterations for the squishy-ball scene with Young's modulus $10^{7}$ (Figure 13 in the main paper). Iterations may terminate early if the residual falls below the tolerance. Results show a nonlinear relationship between the cap and both Newton iterations and total simulation time. 
}
\label{fig:post_coarsening}
\end{figure}

{\color{black}
\paragraph{Post-coarsening Correction.}
To validate the influence of fixed PCG iteration number of our post coarse, we apply a different fixed PCG iterations. This step is not intended to converge to the full-space system, but rather as a corrective mechanism that complements adaptive coarsening: it refines the displacement to compensate for deformation discrepancies among the aggregated fine nodes, thereby enabling the release of additional degrees of freedom in subsequent iterations to avoid the subspace locking problem. We evaluate sensitivity to this step by varying the maximum number of post-coarsening iterations in \autoref{fig:post_coarsening}. In our experiments, 10 iterations, suffice to provide a useful prediction for the aggregator in the next iteration. Increasing to 50 or 100 does not consistently improve this prediction, while very large counts approach the full-space solution at substantially higher cost.

}

\begin{figure}[htbp]
\centering
\includegraphics[width=1.0\columnwidth]{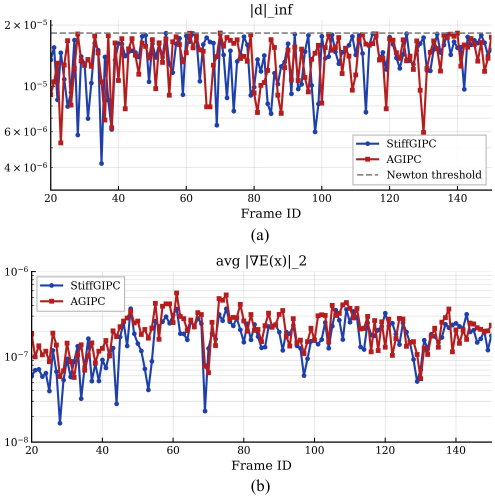}
\caption{\color{black}
\textbf{Analysis of Newton convergence metrics.}
(Top) $L_\infty$ norm of displacement at convergence (dragon scenario with {$\Delta t = 0.01s$}). (Bottom) The corresponding root-mean-square of the energy gradient. 
}
\label{fig:stop_condition}
\end{figure}

{\color{black}
\paragraph{Quantitative Analysis of Convergence Metrics.}

Similar to Trading Spaces \cite{TrustyFLK24}, our Newton solver terminates when the displacement norm satisfies $\|\mathbf{d}\|_\infty/\Delta t \le \varepsilon_d$. In our method, $\mathbf{d}$ denotes the full-space displacement prolongated from the coarse-space and refined by the post-coarsening step, whereas StiffGIPC uses the exact full-space displacement. This implies that our framework converges in the subspace, while StiffGIPC converges in the full-space.
To quantitatively assess the convergence differences between our method and StiffGIPC (see \autoref{fig:stop_condition}), we compare $\|\mathbf{d}\|_\infty$ for each frame across the two frameworks. The results indicate that our $\|\mathbf{d}\|_\infty$ is not necessarily smaller than that of the full-space simulation, allowing the Newton solver to terminate earlier. For an apple-to-apple comparison, we also measure the average full-space gradient norm for both framework. We find that the gradient in our framework also converges to a level similar to that of the full-space simulation. This is due to the combination of adaptive coarsening, affine embedding, and post-coarsening correction, which enables the gradient to diminish to levels comparable to full-space simulation.
This numerical consistency ensures that the resulting dynamics remain visually identical to the reference.

}

\begin{table}[]
	\caption{\color{black}\textbf{Performance analysis of preconditioners.} We compare AGIPC and StiffGIPC with diagonal and MAS preconditioners. \texttt{TotalTime} is the total simulation time. \texttt{Speedup AGIPC/StiffGIPC} gives AGIPC's speedup over StiffGIPC under the same preconditioner. \texttt{Speedup MAS/diagonal} gives the speedup of MAS over the diagonal preconditioner within each framework. The special case in Figure 15 is shown in bold: MAS becomes unstable and thus slower than the diagonal preconditioner. ‘Figure X’ refers to the corresponding figures in the main paper.}
	\label{tab:preconditioner}
	\resizebox{\columnwidth}{!}{%
	\begin{tabular}{c|c|c|c|cc}
	\hline
	\multirow{2}{*}{}                                 & \multirow{2}{*}{preconditioner} & \multirow{2}{*}{framework} & \multirow{2}{*}{\texttt{TotalTime}} & \multicolumn{2}{c}{\texttt{Speedup}}                                        \\ \cline{5-6} 
													  &                                 &                            &                                     & \multicolumn{1}{c|}{\texttt{AGIPC/StiffGIPC}}       & \texttt{MAS/diagonal} \\ \hline
	\multirow{4}{*}{Figure 12 (a)} & \multirow{2}{*}{diagonal}       & StiffGIPC                  & 4496.96                             & \multicolumn{1}{c|}{\multirow{2}{*}{3.25}}          & \multirow{2}{*}{/}    \\ \cline{3-4}
													  &                                 & AGIPC                      & 1384.25                             & \multicolumn{1}{c|}{}                               &                       \\ \cline{2-6} 
													  & \multirow{2}{*}{MAS}            & StiffGIPC                  & 2821.50                             & \multicolumn{1}{c|}{\multirow{2}{*}{3.15}}          & 1.59                  \\ \cline{3-4} \cline{6-6} 
													  &                                 & AGIPC                      & 894.888                             & \multicolumn{1}{c|}{}                               & 1.55                  \\ \hline
	\multirow{4}{*}{Figure 12 (b)} & \multirow{2}{*}{diagonal}       & StiffGIPC                  & 1182.63                             & \multicolumn{1}{c|}{\multirow{2}{*}{2.58}}          & \multirow{2}{*}{/}    \\ \cline{3-4}
													  &                                 & AGIPC                      & 457.887                             & \multicolumn{1}{c|}{}                               &                       \\ \cline{2-6} 
													  & \multirow{2}{*}{MAS}            & StiffGIPC                  & 694.469                             & \multicolumn{1}{c|}{\multirow{2}{*}{2.24}}          & 1.70                  \\ \cline{3-4} \cline{6-6} 
													  &                                 & AGIPC                      & 310.145                             & \multicolumn{1}{c|}{}                               & 1.48                  \\ \hline
	\multirow{4}{*}{Figure 9}            & \multirow{2}{*}{diagonal}       & StiffGIPC                  & 302.674                             & \multicolumn{1}{c|}{\multirow{2}{*}{1.44}}          & \multirow{2}{*}{/}    \\ \cline{3-4}
													  &                                 & AGIPC                      & 210.162                             & \multicolumn{1}{c|}{}                               &                       \\ \cline{2-6} 
													  & \multirow{2}{*}{MAS}            & StiffGIPC                  & 280.418                             & \multicolumn{1}{c|}{\multirow{2}{*}{1.41}}          & 1.08                  \\ \cline{3-4} \cline{6-6} 
													  &                                 & AGIPC                      & 198.735                             & \multicolumn{1}{c|}{}                               & 1.06                  \\ \hline
	\multirow{4}{*}{Figure 10}              & \multirow{2}{*}{diagonal}       & StiffGIPC                  & 10284.5                             & \multicolumn{1}{c|}{\multirow{2}{*}{2.87}}          & \multirow{2}{*}{/}    \\ \cline{3-4}
													  &                                 & AGIPC                      & 3579.15                             & \multicolumn{1}{c|}{}                               &                       \\ \cline{2-6} 
													  & \multirow{2}{*}{MAS}            & StiffGIPC                  & 43601.4                             & \multicolumn{1}{c|}{\multirow{2}{*}{\textbf{8.00}}} & \textbf{0.24}         \\ \cline{3-4} \cline{6-6} 
													  &                                 & AGIPC                      & 5447.82                             & \multicolumn{1}{c|}{}                               & \textbf{0.66}         \\ \hline
	\end{tabular}%
	}
\end{table}

{\color{black}
\paragraph{Analysis of Corner Cases for Preconditioners.}
In most scenarios, MAS preconditioner effectively reduces the total simulation time (see Figure 12 in \autoref{tab:preconditioner}). 
However, there are notable exceptions. In the dolphin example (see Figure 9 in \autoref{tab:preconditioner}), the speedup provided by MAS is less pronounced. In this case, a relatively small Young’s modulus ($10^{4}$) yields a well-conditioned Hessian, for which the diagonal preconditioner already performs reasonably well. Moreover, the large global deformation reduces spatial and temporal coherence, limiting aggregation opportunities and resulting in only a $1.4\times$ speedup. In the twisting example (see Figure 10 in \autoref{tab:preconditioner}), MAS performs even worse than the diagonal preconditioner. Here, extreme deformation in a large stiff model produces a severely ill-conditioned Hessian; multilevel aggregation amplifies numerical instability, leading to extreme large matrix entries and failure to reliably invert the coarse-level blocks. 

In our framework, the numerical instability of MAS is naturally less pronounced. The {\color{black}post-coarsening} step in our framework, which employs a diagonal preconditioner, allows subsequent CG iterations to proceed stably, relieving the conditioning issues introduced by MAS and yielding an $8\times$ speedup in cases where MAS fails in StiffGIPC. However, this comparison may be unfair. To ensure a fair comparison, we configured both StiffGIPC and our framework with the same diagonal block preconditioner, under which StiffGIPC also runs stably. In this setting, our method remains robust and still achieves a $2.87\times$ speedup while preserving visual fidelity.
}

\begin{figure}[htbp]
\centering
\includegraphics[width=\columnwidth]{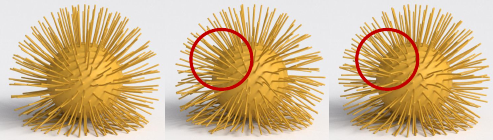}
\caption{\color{black}
\textbf{Different adaptive criteria.}
We evaluate the deformation gradient rate criterion against our Green strain increment criterion on the 1e7 squishy ball example. (Left) Green strain increment criterion. (Middle) The deformation gradient rate criterion, tuned to the best threshold we could find for performance, exhibits artificial stiffening in spiky regions. (Right) Deformation gradient rate criterion with a stricter threshold reduces these artifacts but introduces noticeable temporal lag.
}
\label{fig:deformation_rate}
\end{figure}

{\color{black}
\paragraph{Alternative Adaptive Criteria.}
We introduce affine embedding (Equation 4 in main paper) assigning 12 DoF per coarse node. However, the adaptive coarsening criterion (Equation 3 in main paper) relies on Green strain increment, which mainly distinguishes rigid motions, suggesting an affine-aware criterion may better exploit the embedding. To test this, we adopt the deformation gradient rate approach of \citet{mercier2026affinification} ($\dot{F}=B\dot{x}$) and evaluate it on the stiff ($10^7$ Pa) squishy ball scenario (Figure 13 in the main paper). However, in our experiments, this deformation gradient rate criterion does not outperform our strain increment criterion. As shown in \autoref{fig:deformation_rate}, the deformation-rate results deviate noticeably from the full-space reference, while ours remain visually indistinguishable. Quantitatively (\autoref{tab:deformation_rate}), the deformation gradient rate enables more aggressive coarsening, but the reduction in node count is offset by a sharp increase in Newton iterations, indicating that overly aggressive coarsening compromises convergence and physical fidelity.
}

\begin{table}[]
\caption{\color{black}\textbf{Performance comparison between coarsening criteria.} While the deformation gradient rate criterion enables significantly more aggressive coarsening (resulting in fewer 3DoF/12DoF coarse nodes), it suffers from poor convergence of the Newton solver. The consequent increase in total iterations leads to substantial computational overhead, making it slower overall than our approach.}
\label{tab:deformation_rate}
\resizebox{\columnwidth}{!}{%
\begin{tabular}{c|c|c|c|c}
\hline
									& \texttt{TotalTime} & \texttt{TotalNewtonIter} & 3DoF node & 12DoF node \\ \hline
Green Strain Increment (Ours)                  & 148.417    & 2736              & 8473      & 145        \\ \hline
Deformation Deformation Rate (Optimal) & 206.872    & 10629             & 38        & 45         \\ \hline
Deformation Deformation Rate (Stricter) & 488.308    & 13676             & 1077      & 122        \\ \hline
\end{tabular}%
}
\end{table}

\paragraph{AmgX Configuration.}

\begin{figure}[htbp]
\centering
\includegraphics[width=\columnwidth]{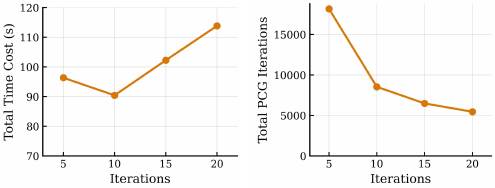}
\caption{
\textbf{AmgX iterations when used as a PCG preconditioner.}
We evaluate the effect of the number of AmgX multigrid V-cycles when used as a PCG preconditioner. The x-axis shows the maximum number of V-cycles per preconditioning step. (Left) Total computation time for 100 frames of the armadillo scene. (Right) Total number of PCG iterations. Increasing the number of V-cycles improves PCG convergence, but incurs higher preconditioning cost.
}
\label{fig:amgx_pcg_iters}
\end{figure}

Here we provide additional details for our comparison with AmgX. We use the \texttt{AGGREGATION} algorithm with a \texttt{SIZE\_4} selector and allow up to 50 multigrid levels. For the smoother, we adopt the symmetric \texttt{MULTICOLOR\_GS} scheme with one presweep and one postsweep. Following the default configuration of the official AmgX library, we set the maximum uncolored percentage to $0.15$ and use the \texttt{MIN\_MAX} matrix coloring scheme. A dense LU solver is employed on the coarsest level. When AmgX is used as a preconditioner for PCG, the number of multigrid V-cycles per preconditioning step has a significant impact on overall performance. We study this effect in \autoref{fig:amgx_pcg_iters}. Using too few V-cycles can degrade PCG convergence, while using too many V-cycles improves the accuracy of the preconditioner but increases the preconditioning cost, potentially slowing down the overall solve. Based on this trade-off, we set the maximum number of V-cycles to 10 in our experiment, which provides a good balance between convergence rate and computational cost.

\section{Mass Distribution and Numerical Stability}

{\color{black}
From a theoretical perspective, aggressively aggregating degrees of freedom while preserving fine resolution at neighboring nodes can lead to severe mass-ratio disparities, potentially causing ill-conditioning in the linear systems arising from Newton iterations. However, across all our experiments, we observe no such numerical instabilities, even when the ratio between the largest and smallest aggregated masses exceeds $10^5$.

This robustness stems from three complementary factors rooted in the structure of our solver.
First, the aggregation induces a Galerkin projection of the original Newton system, yielding a reduced operator $\mathbf{H}_c = \mathbf{U} \mathbf{H}_f \mathbf{U}^T$. This construction preserves the energy geometry of the original system, meaning that the spectral properties relevant to Krylov solvers are inherited by the coarse system. Consequently, large variations in the physical mass represented by individual coarse degrees of freedom do not directly translate into poor conditioning.

Second, the adaptive aggregation naturally aligns with the near-nullspace of the Newton operator. In implicit time integration, the modes that challenge PCG correspond to spatially smooth displacement fields. Because our coarsening criteria group nodes undergoing coherent deformation, these smooth modes remain well represented in the reduced space—a key requirement for stable multigrid-style coarsening.

\begin{figure}[htbp]
\centering
\includegraphics[width=1.0\columnwidth]{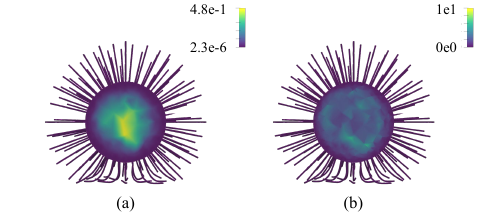}
\caption{\color{black}
\textbf{Localized mass-ratio analysis.}
We evaluate the mass distribution during the Newton iteration exhibiting the highest dynamic range. (a) Global mass distribution. (b) Mass gradient. Despite a maximum-to-minimum mass ratio exceeding $10^5$, the distribution follows a continuous spatial trend. 
}
\label{fig:mass}
\end{figure}

Third, the aggregated mass field exhibits a remarkably smooth spatial gradient, even in the most extreme cases. As shown in \autoref{fig:mass}, although the global mass distribution spans several orders of magnitude, no abrupt mass discontinuities appear between neighboring nodes. This smoothness prevents artificial spectral discontinuities that would otherwise degrade the conditioning of the projected system.

Together, these factors explain why large mass-ratio disparities produce no numerical artifacts in practice. While additional techniques such as localized mass scaling or hierarchical averaging could further formalize this behavior, our empirical results indicate that the current adaptive strategy already preserves the spectral structure necessary for robust PCG convergence.

}

        
	\bibliographystyle{ACM-Reference-Format}
	\bibliography{ref}
\clearpage